\documentclass[10pt,twocolumn,notitlepage,floatfix,footinbib,superscriptaddress,showkeys,longbibliography,aps]{revtex4-1}

\usepackage{dcolumn}
\usepackage{amsfonts}
\usepackage[pdftex,dvipsnames]{xcolor}
\usepackage{array}
\usepackage{mathtools}
\usepackage{booktabs}
\usepackage{xspace}
\usepackage[bookmarks=false,linkcolor=blue,urlcolor=blue,colorlinks,citecolor=blue]{hyperref}

\newcolumntype{M}[1]{>{\centering\arraybackslash}m{#1}}

\usepackage[colorinlistoftodos, prependcaption, textsize=normal]{todonotes}
\definecolor{amethyst}{rgb}{0.6, 0.4, 0.8}
\definecolor{blueish}{RGB}{5, 95, 139}

\newcommand{\deltaind}{$\Delta_{\rm ind}$\xspace}
\newcommand{\vg}{$V_g$\xspace}
\newcommand{\gf}{$g$\xspace}
\newcommand{\ps}{SP\xspace}   % To abbreviate Schroedinger Poisson as PS or SP
\newcommand{\gbare}{g^{\text{bare}}}

\newcommand{\Pf}{{\rm Pf}}
\newcommand{\eps}    {\epsilon}
\DeclareMathOperator {\sgn}{sgn}
\newcommand{\ceq}[1] {(\ref{#1})}
\newcommand{\rr}     {{\bf r}}

\begin{document}

\title{Effects of gate-induced electric fields on semiconductor Majorana nanowires}

\author{Andrey E. Antipov}
\email{andrey.antipov@microsoft.com} 
\affiliation{Station Q, Microsoft Research, Santa Barbara, California 93106-6105, USA}
\author{Arno Bargerbos}%
\affiliation{QuTech, Delft University of Technology, 2600 GA Delft, The Netherlands}
\affiliation{Kavli Institute of Nanoscience, Delft University of Technology, 2600 GA Delft, The Netherlands}
\author{Georg W. Winkler}
\affiliation{Station Q, Microsoft Research, Santa Barbara, California 93106-6105, USA}
\author{Bela Bauer}
\affiliation{Station Q, Microsoft Research, Santa Barbara, California 93106-6105, USA}
\author{Enrico Rossi}%
\affiliation{Department of Physics, William \& Mary, Williamsburg, VA 23187, USA}
\affiliation{Station Q, Microsoft Research, Santa Barbara, California 93106-6105, USA}
\author{Roman M. Lutchyn}
\affiliation{Station Q, Microsoft Research, Santa Barbara, California 93106-6105, USA}

\date{\today}

\begin{abstract}

We study the effect of gate-induced electric fields on the properties of semiconductor-superconductor hybrid nanowires which represent a promising platform for realizing topological superconductivity and Majorana zero modes. Using a self-consistent Schr\"odinger-Poisson approach that describes the semiconductor and the superconductor on equal footing, we are able to access the strong tunneling regime and identify the impact of an applied gate voltage on the coupling between semiconductor and superconductor. We discuss how physical parameters such as the induced superconducting gap and Land\'e g-factor in the semiconductor are modified by redistributing the density of states across the interface upon application of an external gate voltage. Finally, we map out the topological phase diagram as a function of magnetic field and gate voltage for InAs/Al nanowires.

\end{abstract}

\maketitle

\section{Introduction} 

Composite heterostructures provide an opportunity to realize exotic phases of matter by exploiting the properties of individual components. A particularly interesting example involves semiconductor-superconductor hybrid structures which represent a promising platform for the realization of topological superconductivity~\cite{Beenakker13a, Alicea12a, Leijnse12, Stanescu13b, Franz15, DasSarma15, MasatoshiSato2016, Aguado2017,Lutchyn2018}. Topological superconductors support exotic neutral excitations consisting of an equal superposition of an electron and a hole -- Majorana zero-energy modes (MZMs)~\cite{Read00,Ivanov01,Kitaev01}. Due to the particle-hole symmetry in a superconductor, such modes appear at zero energy and, thus, there is no cost to occupy these states. This leads to a growing degeneracy of the ground state as the number of MZMs is increased, a hallmark of topological superconductors. Theory predicts that exchanging the position of MZMs~\cite{Moore91,Read00} or performing certain non-local measurements of the charge encoded in a collection of MZMs~\cite{Bonderson08b} leads to a nontrivial transformation within the degenerate ground-state manifold, and represents a non-Abelian operation which is independent of the details of its execution. This property of topological superconductors has generated a lot of excitement in the condensed matter physics, quantum information, and material science communities~\cite{Wilczek2009,Stern10,Brouwer_Science,Lee14} as it opens up the possibility of Majorana-based topological quantum computing~\cite{kitaev2003fault,Nayak08,DasSarma15,Lutchyn2018}.

Realizing topological superconductivity in the laboratory is not an easy task since the originally proposed models~\cite{Read00,Kitaev01} involved spinless p-wave superconductivity. Electrons in solids have spin-$\frac{1}{2}$ and most of the common superconductors have s-wave pairing which involves electrons with opposite spins. Therefore, quenching spin degeneracy and preserving superconducting pairing is quite non-trivial. One way to overcome the problem is to use materials with a strong spin-orbit interaction which couples spin and orbital degrees of freedom. A number of platforms for realizing MZMs in the laboratory have been recently proposed~\cite{Fu08, Fu&Kane09, Cook11, Sun2016, Sau2010, Alicea10, Lutchyn10, Oreg10, SukBum2011, Duckheim2011, Potter12, Choy11,Martin12,nadj2013, klinovaja2013,braunecker2013,vazifeh2013,pientka2013, Nakosai2013, Kim14, brydon2014, Li14, Kotetes2014, Ojanen2015, Nadj-Perge14,Ruby15,Pawlak16, JZhang16}. The most promising proposal for realizing MZMs is based on one-dimensional (1D) semiconductor-superconductor (SM-SC) hybrid structures~\cite{Lutchyn10, Oreg10} and involves a semiconductor with strong spin-orbit coupling (such as InAs or InSb) and an s-wave superconductor (such as Al). In this proposal, a magnetic field or another time-reversal breaking perturbation is needed to drive the system into the spinless topological regime~\cite{Lutchyn10, Oreg10}. This proposal has triggered significant experimental activity~\cite{Mourik12,Rokhinson12,Deng12,Churchill13,Das12,Finck12,Chang14,Krogstrup2015,Albrecht16,Zhang16,Chen2016, Deng2016, Suominen2017, Nichele2017,Gazibegovic2017, Zhang2017, Zhang2017a, Sestoft2017,Deng2017, Sole2017, Geresdi2017}, and there is a 
compelling body of experimental evidence that MZMs have been realized in these systems. For a very recent example, see Ref.~\cite{Zhang2017} which reports a robust quantized $2e^2/h$ zero-bias conductance consistent with the Majorana scenario. 

\begin{figure}
    \begin{center}\includegraphics[width=\columnwidth]{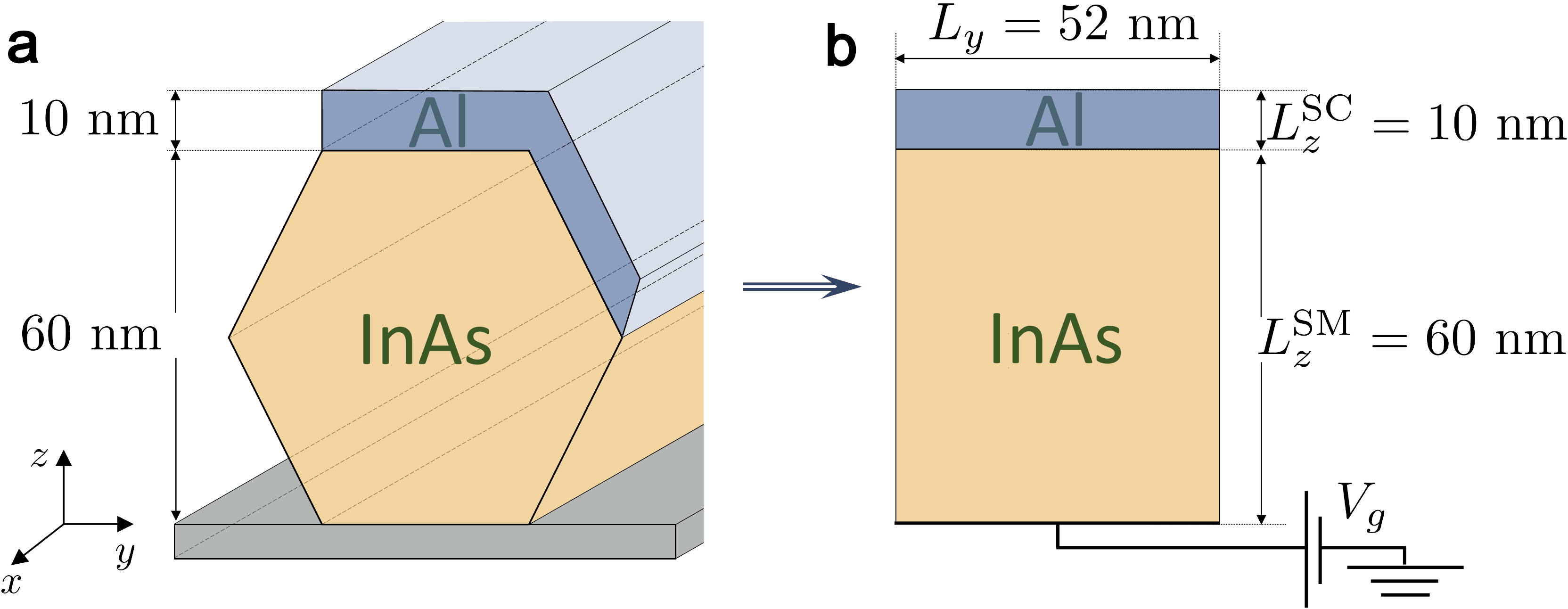}\end{center}
    \caption{(Color online) (a) SM-SC heterostructure based on hexagonal nanowire. $10$ nm thick Al layer (blue) is deposited on 2 facets of InAs (brown) hexagonal wire with a height of $60$ nm. The back gate is shown schematically in gray. (b) Rectangular geometry of the wire that supports the same number of subbands. The back gate is emulated by a boundary condition at the bottom.}
    \label{fig:fig1}
\end{figure}

Much of the progress in realizing MZMs with proximitized nanowires is attributed to the material science advance in fabricating semiconductor-superconductor heterostructures. In the first generation of experiments~\cite{Mourik12,Rokhinson12,Deng12,Churchill13,Das12,Finck12} the superconductor was deposited ex-situ which required removing the native oxide forming on the semiconductor's surface due to air exposure. In the second generation of experiments the thin aluminum shell~\cite{Krogstrup2015} is deposited epitaxially and is thus grown on pristine SM facets without breaking the vacuum, see Fig.~\ref{fig:fig1}. Tunneling spectroscopy measurements of the induced superconducting gap~\cite{Chang14, Deng2016, Nichele2017, Zhang2017, Sole2017} in such samples exhibit a large induced gap (i.e. close to the bulk gap of the superconductor) which indicates that the improved epitaxial interfaces are characterized by a strong hybridization of the states in the semiconductor and superconductor. In this strong tunneling regime, many physical parameters such as the g-factor and spin-orbit coupling are strongly renormalized due to the hybridization. In order to quantitatively understand the hybridization and its implications on the band structure as well as other physical properties, one has to consider the band offset at the superconductor-semiconductor interface.  Depending on the sign of the band offset one can have either a Schottky barrier or an accumulation layer~\cite{tung2001,luth2011, abe2002,feng2016}. Based on preliminary ARPES studies~\cite{ARPES}, one finds that the band offset for epitaxially grown InAs/Al heterostructures is $-(200-300)$meV supporting the accumulation layer scenario. 

Proper theoretical treatment of the strong coupling regime is also necessary to understand
how external gates affect the electronic state, and in particular the topological nature, of SM-SC heterostructures.
Furthermore, recent proposals for realizing scalable architectures for topological quantum computation with MZMs rely on fine electrostatic control~\cite{Landau16,Vijay2016, Plugge16a,Karzig2016,plugge2017}. Thus, understanding the effect of electric fields on the low-energy properties of the proximitized nanowires is critical both for the interpretation of the existing Majorana experiments~\cite{Deng2016, Nichele2017, Zhang2017, Albrecht16,Sole2017} as well as for the optimization of proposed Majorana devices~\cite{Lutchyn2018}. 

In order to understand the physical properties of the proximitized nanowires, one needs to solve the electrostatic and quantum-mechanical problems self-consistently, i.e. perform Schr\"odinger-Poisson (\ps) calculations. Compared to the case of purely semiconducting heterostructures~\cite{Stern1972,Ando1982, stern1984}, the problem at hand is much more challenging technically because it involves disparate materials with very different effective masses, Fermi energies, $g$-factors etc. (see Table~\ref{table:params}). In other words, the standard numerical tools based on the continuum mass approximation cannot be applied to semiconductor-superconductor hydrid systems. Therefore, modeling of the semiconductor-superconductor hybrid structures requires developing numerical techniques which can effectively take into account different length scales in the semiconductor and superconductor. 

Previous effective models for superconductor-semiconductor hybrids~\cite{Stanescu2011, Prada12, Rainis13, Cole2015, Reeg17, Sticlet2017} do not properly describe the experimental system and provide only qualitative predictions for the electric field dependence.  These models rely on independent phenomenological parameters such effective masses, spin-orbit couplings, g-factors as well as tunneling strength between semiconductor and superconductor.
While this approach may be suitable for the weak tunneling regime,
naive extensions of such models to the strong coupling limit are inadequate. This is because the electric field applied to the semiconductor can drastically change the electrons' confinement, i.e. push or pull electron density in the semiconductor to or away from the interface. This in turn strongly affects physical parameters of the system, including, as we will see, the tunneling rate, effective spin-orbit coupling, $g$-factor as well as induced superconducting gap.

More advanced models have been introduced recently~\cite{Soluyanov16, Vuik16, Degtyarev2017, Dominguez2017, Winkler2017} which treat the effects of an electric field within some effective models where the superconductor is taken into account via boundary conditions. This approach, while being computationally advantageous, does not take into account the effects arising from the redistribution of the wavefunction between the semiconductor and the superconductor. In this work, we treat the superconducting (SC) and semiconducting (SM) degrees of freedom explicitly on the same footing. Using an adaptive discretization algorithm for the SM and SC components, we develop an effective model which is computationally tractable and allows us to adequately capture the effect of the gate-induced electric field on the heterostructure. 

Our results allow one to understand and interpret recent experiments investigating the electric field and disorder dependence of the effective parameters~\cite{Deng2016, Nichele2017, Zhang2017, Albrecht16, Sole2017}. 
They have motivated the recent systematic experimental study of the tunability of the superconductor-semiconductor coupling in Majorana
nanowires~\cite{deMoor2018}. 
A better understanding of the effect of the external electric fields on the phase diagram of superconductor-semiconductor nanowires is important to reinforce the confidence that the observed zero-bias peaks are due to Majorana zero modes.
Our results describe the conditions necessary to achieve a strong tunneling regime between superconductor and semiconductor and show how external electric fields and the presence of disorder affect the topological phase diagram of SC-SM nanowires. 
They provide an important contribution for the realization of the next-generation experiments designed to verify the non-Abelian nature of the modes responsible for the zero-bias peak observed so far in transport measurements and to give an indisputable proof of the realization of Majoranas.

The paper is organized as follows. We begin with a discussion of the Setup and Methods in Sec.~\ref{sec:setup} where we provide technical details of the Schr\"odinger-Poisson approach. In Sec.~\ref{sec:results} we present our results. We first focus on the limit of zero magnetic field and then discuss the behavior at finite magnetic fields. We then discuss the resulting topological phase diagram. We conclude the section with studying the effects of disorder that change the strength of SM-SC coupling. We summarize our results in Sec.~\ref{sec:conclusions} and discuss their relevance for current and future experiments.

%======

\section{Setup and Methods} \label{sec:setup}

We consider the system shown in Fig.~\ref{fig:fig1}. 
The nanowires used in current experiments typically have a hexagonal shape as shown in Fig.~\ref{fig:fig1}~(a). 
The cross section of the wire, which we take to be the $(y,z)$-plane, consists of a $10$~nm thick Al film (blue) covering 2 facets of InAs nanowire (orange). The electrostatic environment is controlled by a back-gate (gray). For practical reasons we do not explicitly treat this gate and the separating dielectric medium in our calculations, but rather take the gate into account only as a boundary condition for the potential in the wire. In order to convert this into the actual
voltage applied to the gate (which is sample-dependent), the distance to the gate and the dielectric constant have to be taken into account.     For the devices of interest, the length of the wire,  $L_x$, is much larger than its transverse dimensions $L_{y,z}$.  

The presence of the Al layer breaks the hexagonal symmetry of the nanowire cross section and, as shown in Fig.~\ref{fig:fig2}, causes
the formation of an electrostatic potential that strongly confines the electrons close to the SM/Al interface.
For this reason the hexagonal cross section of the wire can be well approximated by an effective rectangular cross section,
as shown in Fig.~\ref{fig:fig1}~(b). 
We will henceforth refer to the effective wire with rectangular cross section 
as the \textit{slab model}. 
By choosing $L_y$ for the slab model to be such that the number of cross sectional modes is the same as for
the hexagonal cross section wire, the use of the slab model does not cause any significant loss of accuracy
and significantly simplifies the numerical implementation and solution of the \ps problem.

\begin{table}[t]
\begin{center}
\begin{tabular}{|p{0.4\columnwidth}|M{0.25\columnwidth}|M{0.25\columnwidth}|}
\specialrule{1.5pt}{1pt}{1pt}
{\bf Parameter} & {\bf InAs} & {\bf Al} \\
\specialrule{1.5pt}{1pt}{1pt}
$m^{*}$ & $0.026$~\cite{Vurgaftman2001} & $1$ \\
$\epsilon_r$ & $15.15$ & \\
$W$, eV & $-0.25$~\cite{ARPES} & \\
\hline
$\gbare$ & $-15$~\cite{Pidgeon1967} & $2$ \\
$\alpha$, eV$\cdot$ nm  & $0.01$~\cite{Weperen2015} & $0$ \\
$ \varepsilon_F$, eV & $0$ & $11.27$~\cite{Cochran1958} \\  
$\Delta_0$, meV & $0$ & $0.34$~\cite{Cochran1958} \\ 
\hline
$L_z$, nm & $60$ & $10$ \\
$L_y$, nm & $52$ & $52$ \\
\hline
\end{tabular}
\end{center}
\caption{Physical parameters for InAs and Al.}\label{table:params}
\end{table}

The Hamiltonian for the heterostructure in the normal state can be written as ($\hbar=1$)
\begin{align}\label{eq:H_n}
&\hat H_n = -\partial_z \left[\frac{1}{2m^*(z)} \partial_z \right] + \frac{1}{2m^*(z)} \left(\hat k_x^2 + \hat k_y^2\right) \\ 
& \!+\!\phi(z)\!-\! \varepsilon_F(z) \!-\! \alpha(z)\left(\hat k_x \sigma_y \!-\! \hat k_y \sigma_x\right) \!+\! \frac{\mu_B \gbare(z) B}{2} \sigma_x,\nonumber
\end{align}
where the spatially-dependent effective mass $m^*(z)$, Fermi energy $\varepsilon_F(z)$, spin-orbit coupling strength $\alpha(z)$,
and \gf factor $g(z)$ are equal to $m^{*}(z)=m_{SM}$ ($m^{*}(z)=m_{SC}$) for $z<60$~nm ($z>60$~nm) and similarly for $\varepsilon_F(z)$, $\alpha(z)$, and $g(z)$; $\hat k_x, \hat k_y$ are the momentum operators in the $x$ and $y$ direction, respectively; 
$\phi(z)$ is the electrostatic potential,
$\sigma_{x,y,z}$ are the Pauli-matrices in spin space,
$\mu_B$ and $B$ are the Bohr magneton
and the external magnetic field, respectively. 
The values for the material parameters used henceforth are given in Table~\ref{table:params}.

In this work we investigate bulk properties of the heterostructure. Therefore, we assume henceforth that the nanowire is infinitely long and translationally invariant along the $x$ direction. This allows one to use as a basis plane waves along the $x$ direction and to replace the operator $\hat k_x$ in 
\eqref{eq:H_n} by its eigenvalue. 
In the clean limit considered here, due to the finite-size quantization in the $y$ and $z$ directions, the spectrum of the system consists of effectively 1D subbands.
We obtain the eigenvalues and eigenstates of the resulting Hamiltonian $\hat H_n(\hat k_x\to k_x)$
corresponding to these subbands
via a mode decomposition in the $y$ direction and by replacing the derivatives with respect to $z$ 
with finite differences using a non-uniform grid~\cite{tan1990} with two different spacings corresponding to the semiconducting and superconducting components, respectively. The spacings are chosen such that $dz < \pi/k_F$ in order to minimize discretization errors. Using a non-uniform spacing significantly alleviates the computational cost and allows us to systematically study the phase diagram of the problem.

In the absence of spin-orbit coupling and disorder the discrete modes along the $y$ direction are
\begin{equation}
\psi_{n_y}^{\alpha = 0}(y)=\sqrt{\frac{2}{L_y}}\sin\left(\frac{\pi n_y}{L_y} y\right) \label{eq:wf_y_sine}
\end{equation} 
with the different $n_y\in \mathbb{N}$ modes being decoupled. 
The spin-orbit coupling term hybridizes them~\cite{Stanescu2011}. The corresponding matrix elements are
\begin{equation}
\begin{aligned}
\mathcal{A}_{n_y n'_y}(z) = & 
-i \alpha(z) \langle \Psi_{n_y}^{\alpha = 0} | \frac{d}{dy} | \Psi_{n'_y}^{\alpha = 0} \rangle \\ 
= & - \frac{2i\alpha(z)}{L_y} (-1 + (-1)^{n_y+n'_y}) \frac{n_y n'_y}{n_y^2 - (n'_y)^2}.
\end{aligned}\label{eq:Amatrix}
\end{equation}
In the $\{\psi_{k_x, n_y,z} = e^{ik_x x}\Psi_{n_y}^{\alpha = 0}(y)\delta(z)\}$ basis the Hamiltonian matrix takes the form
\begin{equation}
H_n = H_n^{\alpha = 0} - \alpha(z) k_x \sigma_y + \mathcal{A}(z) \otimes \sigma_x,
\label{eq:H_ky}
\end{equation}
where $\mathcal{A}$ is the $N_y \times N_y$ matrix with elements given by Eq.~(\ref{eq:Amatrix}), $N_y$ is the number of discrete modes along the $y$ direction considered and $H_n^{\alpha = 0}$ is the matrix obtained by projecting the Hamiltonian (\ref{eq:H_n}) for $\alpha = 0$ on this basis.

We treat the s-wave superconductor at the BCS mean-field level. The Bogoliubov-de-Gennes (BdG) Hamiltonian for the system can be written as 
\begin{multline} 
H = H_n^{\alpha = 0} \otimes \tau_z - \alpha(z) k_x \sigma_y \otimes \tau_z \\
     + \mathcal{A}(z)\otimes \sigma_x \otimes \tau_0 - \Delta(z) \sigma_y  \otimes \tau_y,
\end{multline}
where $\tau_{0,x,y,z}$ are Pauli matrices in Nambu (particle-hole) space.
We include the superconducting pairing only in the superconductor, i.e. $\Delta(z)=\Delta_0$ for $z>60$~nm, where $\Delta_0$
is the SC gap of Al (see Table~\ref{table:params}), and $\Delta(z)=0$ for $z < 60 \mathrm{nm}$.
In a finite magnetic field, the superconducting gap in the Al shell is suppressed due to the inclusion of a finite $g$-factor for the Al (see Table~\ref{table:params}). Given that the Al film is very thin, see Fig.~\ref{fig:fig1}, we neglect orbital effects due to the magnetic field.

\subsection{Electrostatics}

\begin{figure}
\begin{center}\includegraphics[width=0.9\columnwidth]{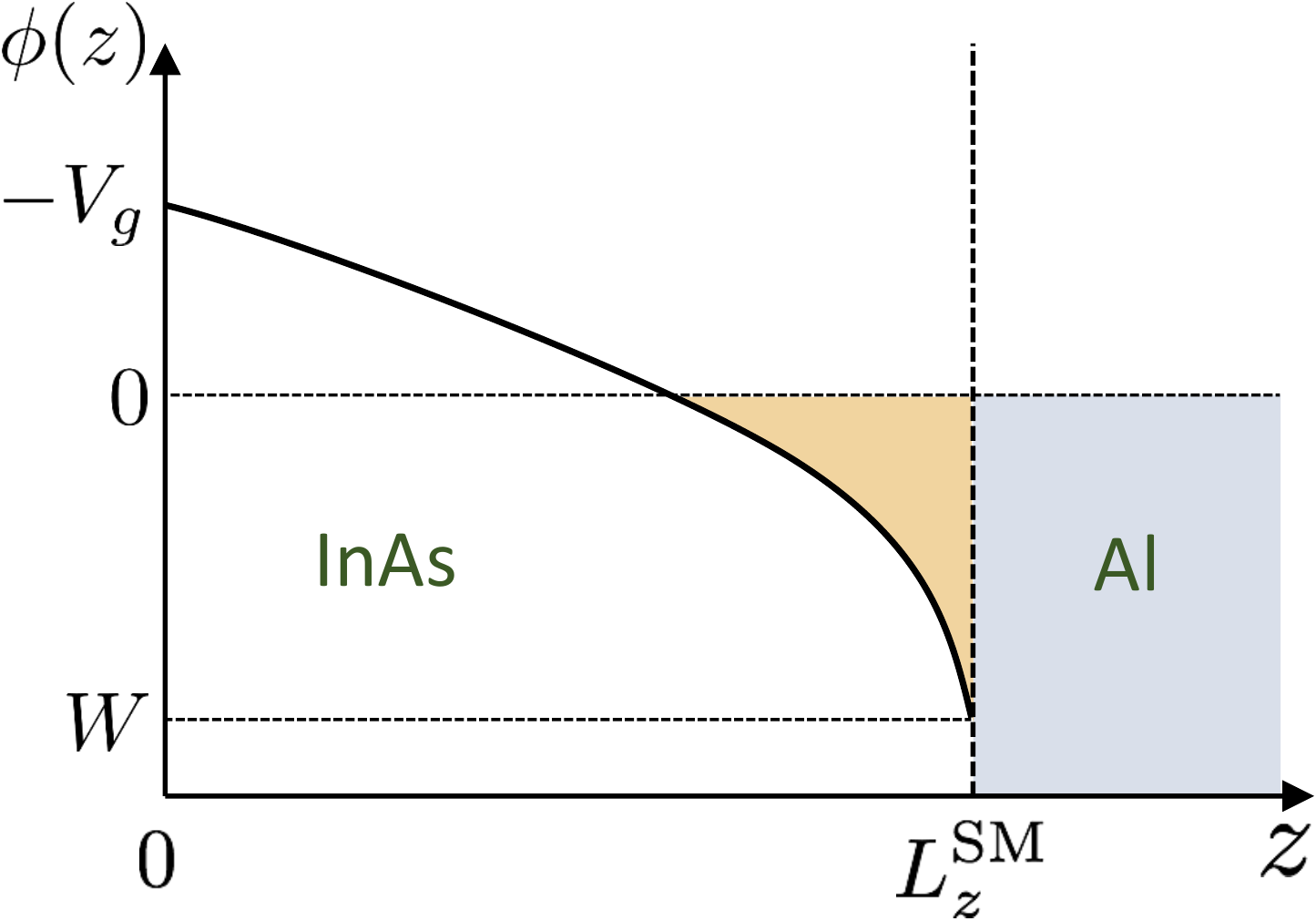}\end{center}
\caption{(Color online) The electrostatic calculation uses Dirichlet boundary conditions at $z=0$ and $z=L_z^{\mathrm{SM}}$, i.e. the top and bottom of the semiconducting wire. At $z=0$, the boundary condition is given by the gate voltage $V_g$, while at the interface to the aluminum it is given by the band offset $W$ (see also Table~\ref{table:params}). This leads to an accumulation layer at the interface.}
\label{fig:fig2}
\end{figure}

In order to obtain the electrostatic profile, one has to solve the \ps equations self-consistently. Given that the BCS mean-field approximation breaks charge conservation this is a non-trivial task, see, e.g., discussion in Ref.~\cite{Knapp2017}. However, electrostatic screening of a metal is only weakly modified by the superconductivity with the small parameter being $\Delta_0/\epsilon_F \ll 1$. 
As a consequence, to obtain the electrostatic potential within this accuracy the charge density entering the 
Poisson equation can be calculated neglecting the superconducting pairing,
i.e. using the Hamiltonian $H_n$ instead of the full Hamiltonian $H$.
The effects of the spin-orbit coupling and Zeeman terms~\cite{Vuik16} on the total electron density profile
$n(z)$ are also very small and can be neglected.
Thus, to solve the full problem we follow a two step approach:
we first solve the \ps problem in the normal state taking $\alpha=0$ and $B=0$ 
to obtain the electrostatic profile.
We then use the obtained electrostatic profile to find the eigenvalues
and the eigenstates of the system for $\Delta_0\neq 0$, $\alpha \neq 0$,
and different values of $B$. 

The first step consists in solving self-consistently the Schr\"odinger equation $H_n|\Psi\rangle = E|\Psi\rangle$,
%xxx
requiring $\Psi$ to vanish at the boundaries of the system,
%xxx
\footnote{Note that the presence of $z$-dependent masses, spin-orbit coupling and potentials does not allow to separate variables in $z$ and other directions and therefore a diagonalization of a Hamiltonian matrix would be required for every different set of momenta.}
and the Poisson equation
\begin{equation}
    \partial^2_z \phi(z) = -\frac{n(z)}{\epsilon_0\epsilon_r},
    \label{eq:poisson}
\end{equation}
where 
$n(z) = (2\pi L_y)^{-1}\sum_{n_y, E[\Psi] <= 0}\int dk_x\left|\Psi_{k_x, n_y}(z)\right|^2$,
$\epsilon_r$ is the relative dielectric constant of the SM, see Tab.~\ref{table:params},
and $\epsilon_0$ is the vacuum dielectric constant.
The setup for the Poisson equation is shown in Fig.~\ref{fig:fig2}. 
At $z=L_z^{\mathrm{SM}}$ the boundary condition for $\phi(z)$ is given by the band offset $W$ between the SM and the SC.  The boundary condition at $z=0$ is set by the back gate. 
The coupled Schr\"odinger-Poisson equations  are solved iteratively until convergence is achieved, using Anderson's mixing algorithm \cite{Anderson1965}.

\subsection{Band structure}

The calculated electrostatic profile $\phi(z)$ is inserted into the full Hamiltonian $H$ to
obtain the band structure $\{\varepsilon^{(n)}(k_x)\}$ and the corresponding eigenstates of the nanowire.
Since the chemical potential is included in the Hamiltonian, the effective Fermi energy for each band is set simply by
the bottom of the band.
We can find the Fermi momentum in each
band, $k_F^{(n)}$, by solving $\varepsilon^{(n)}(k_F^{(n)})=0$. The Fermi velocity is given by
$v_F^n = \frac{d\varepsilon^{(n)}}{d k^{(n)}}|_{k^{(n)}=k_F^{(n)}} \simeq 2|\varepsilon ^{(n)} (k_x = 0)| / k_F^{(n)}$. 
In addition, from the eigenstates at $k=k_F$ we extract how strongly different subbands are coupled to the superconductor, which we define through the weight of the corresponding state in the superconductor 
\begin{equation} \label{eq:Wsc}
W^{\mathrm{SC}} = 1 - \sum_{n_y, \sigma} \int_{0}^{L_z^{\mathrm{SM}}} |\Psi(k_F)|^2 dz
\end{equation}

We define the gap as the minimum of the energy of the first excited state:
$E_g = \min_{k_x} |\varepsilon^{(n)}_{\mathrm{BDG}}(k_x)|$. 
At zero magnetic field, $E_g$ gives an estimate for the induced gap $\Delta = E_g(B=0)$.

\section{Results}\label{sec:results}

In this section we discuss the results of our numerical simulations. We first discuss only the electrostatic
problem for both a model of a hexagonal wire and the slab model introduced above. We then investigate the nature of the electronic states in a limit of strong coupling between the semiconductor and superconductor and discuss their superconducting properties at zero magnetic field. Then we study properties of the hybrid nanowires in a finite magnetic field and  obtain estimates for the effective g-factor in the hybrid structure. We present the topological phase diagram and compare it with previous results~\cite{Lutchyn2011,Stanescu2011}. Finally, we present the results for the wires with the disorder potential present in the superconductor and show its impact on the induced gap and the phase diagram.

\subsection{Electrostatics and density distribution}

\subsubsection{Hexagonal cross section}

In order to obtain the correct number of subbands for a given gate configuration for the wire with the hexagonal cross section it is sufficient to solve the \ps problem using the Thomas Fermi approximation
and simply requiring the wave function to vanish at the boundaries of the cross section.
The solution of the full \ps problem is computationally expensive due to the shape of the cross section and unnecessary for the purpose of simply estimating the number of cross sectional modes. We perform this calculation in COMSOL and obtain eigenstates using the Kwant package~\cite{kwant}.

\begin{figure}
\includegraphics[width=0.9\columnwidth]{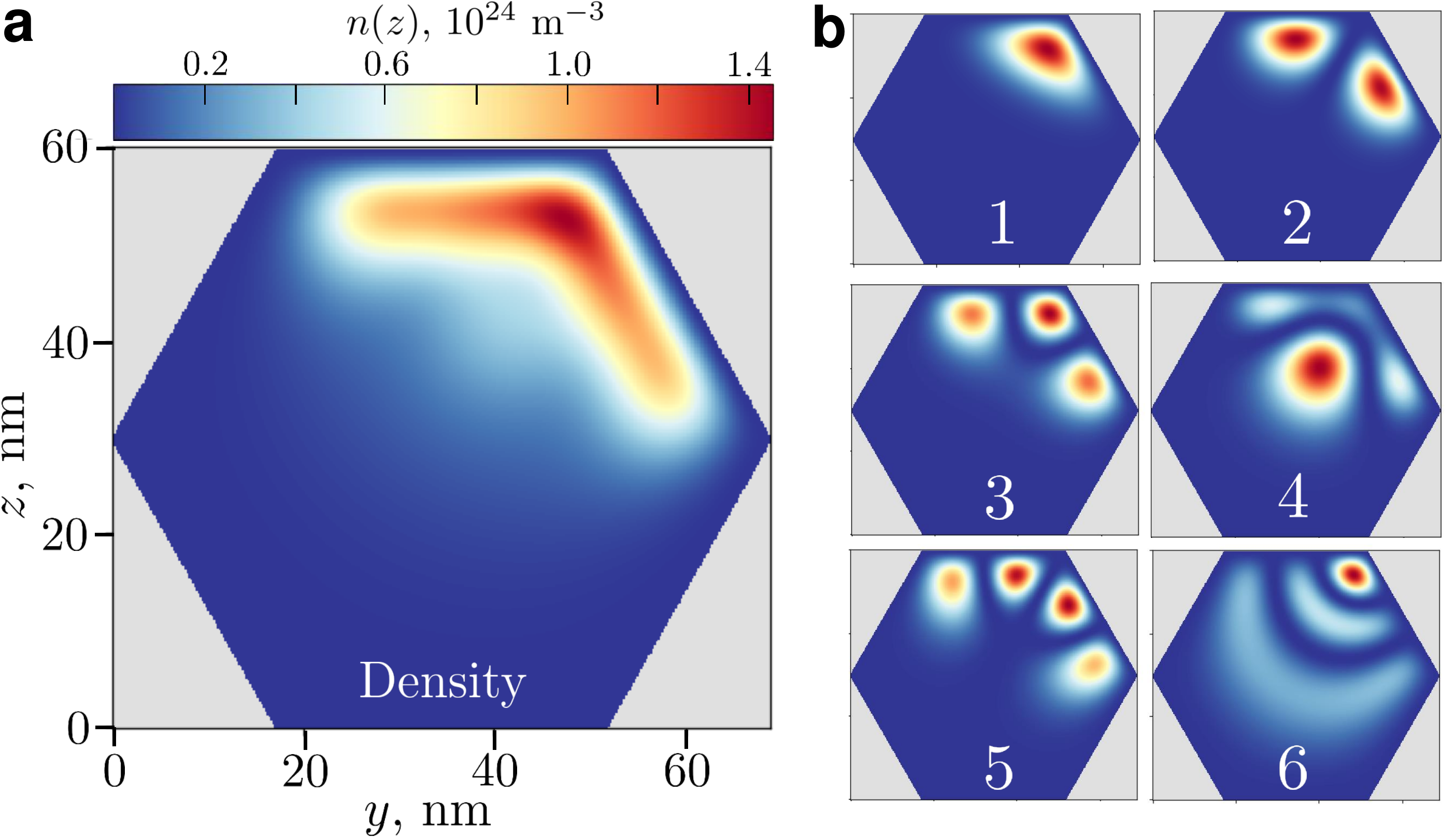}
\caption{(Color online) (a) Electronic density in the cross-sectional cut of the nanowire for $V_g=0$ obtained the Thomas Fermi approximation. (b) Square modulus of eigenstates of the wire in the normal state at $B=0$ with energies $-0.096$, $-0.068$, $-0.052$, $-0.023$, $-0.021$, and $-0.006$ eV for panels 1 to 6 respectively.}
\label{fig:fig3}
\end{figure}

Our results are summarized in Fig.~\ref{fig:fig3},
where we show the density for all occupied modes below the Fermi energy. This calculation does not explicitly treat
the aluminum shell; instead, it assumes that the only effect of the presence of the Al layer is to induce a band offset.
We set this band offset to $W=-0.25~\mathrm{eV}$~\cite{ARPES}, see Table~\ref{table:params}.
The approximations used to obtain the results of Fig~\ref{fig:fig3} cause quantitative inaccuracies for the local density of states (LDOS) and the carrier density profile. 
However, these results are sufficiently accurate to estimate the number of electronic cross-sectional modes below the Fermi energy for a given $V_g$.
In addition, the results of Fig.~\ref{fig:fig3} (a)~show the qualitatively correct result
that for $V_g\leq 0$
most of the charge density is localized at the semiconductor (SM)-superconductor (SC) interface 
due to the strong band offset between the InAs and Al. 
This fact means that for the slab model, the thickness of the SM wire in the $z$ direction does not affect the electronic properties in a significant way as long as it is few times larger than the confinement length in the $z$ direction ($\sim 20$~nm). 
The effective width $L_y$ of the slab model can then be fixed by requiring the number of subbands to be equal to the number of cross-sectional modes obtained from
the hexagonal calculation, as long as $L_y$ is also larger than the confinement length in the $z$ direction.
For $V_g=0$ the hexagonal cross section results show that there are six modes, see Fig.~\ref{fig:fig3}~(b).
From this we obtain that for the slab model $L_y=52$~nm, larger than the confinement length for $V_g=0$.
In the remainder all the results are obtained using the effective slab model with $L_y =52$~nm width and $L_z^{\mathrm{SM}} = 60$~nm thickness for the SM
and $L_z^{(\rm SC)}=10$~nm for Al, as shown in Table~\ref{table:params}.

\subsubsection{Slab model}

\begin{figure}[!b]
\includegraphics[width=0.96\columnwidth]{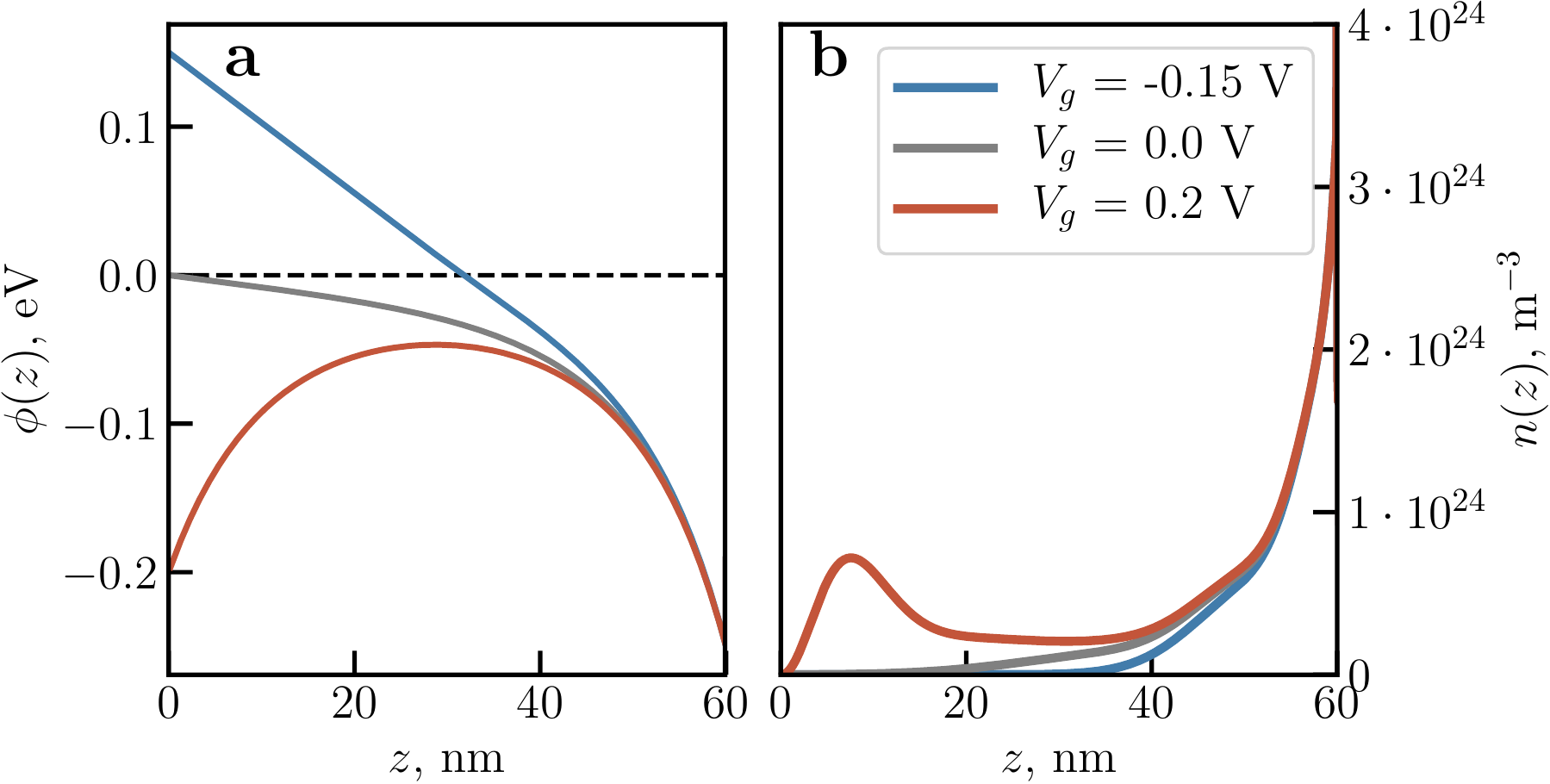}
\caption{(Color online) (a) Electrostatic potential profile $\phi(z)$ and (b) electronic density $n(z)$ in the semiconducting part of the system obtained from a self-consistent Schr\"odinger-Poisson calculation for three representative values of the gate voltage.} 
\label{fig:fig4}
\end{figure}

We now switch to the slab model, which explicitly treats the superconducting Al shell. We self-consistently solve the coupled Schr\"odinger-Poisson (SP) equations for three different values of $V_g$ to obtain
the electrostatic potential $\phi(z)$ and the density $n(z)$, respectively shown in panels (a) and (b) of
Fig.~\ref{fig:fig4}. Since the Al shell is taken to be metallic with an extremely
short screening length, the electrostatic potential is assumed to be constant throughout the Al.
The dashed line in Fig.~\ref{fig:fig4}~(a) shows the Fermi level in Al.
It is worth pointing out that because $\Delta_0 \ll \epsilon_F$, including the pairing term for the Al
makes only a negligible difference to the electrostatic profile. 

For $V_g\leq 0$ the electrostatic potential confines the carrier density in a layer about 20~nm wide close to the SM/Al interface, as
shown in Fig.~\ref{fig:fig4}~(b).
For $V_g>0$ the electrostatic potential is below the Fermi energy also on the gate side. This allows the accumulation
of charges also near the gate, as shown by the result in Fig.~\ref{fig:fig4}~(b)
for $V_g=0.2~V$.

\subsection{Nature of electronic states in strong-coupling limit}

\begin{figure}
\includegraphics[width=0.96\columnwidth]{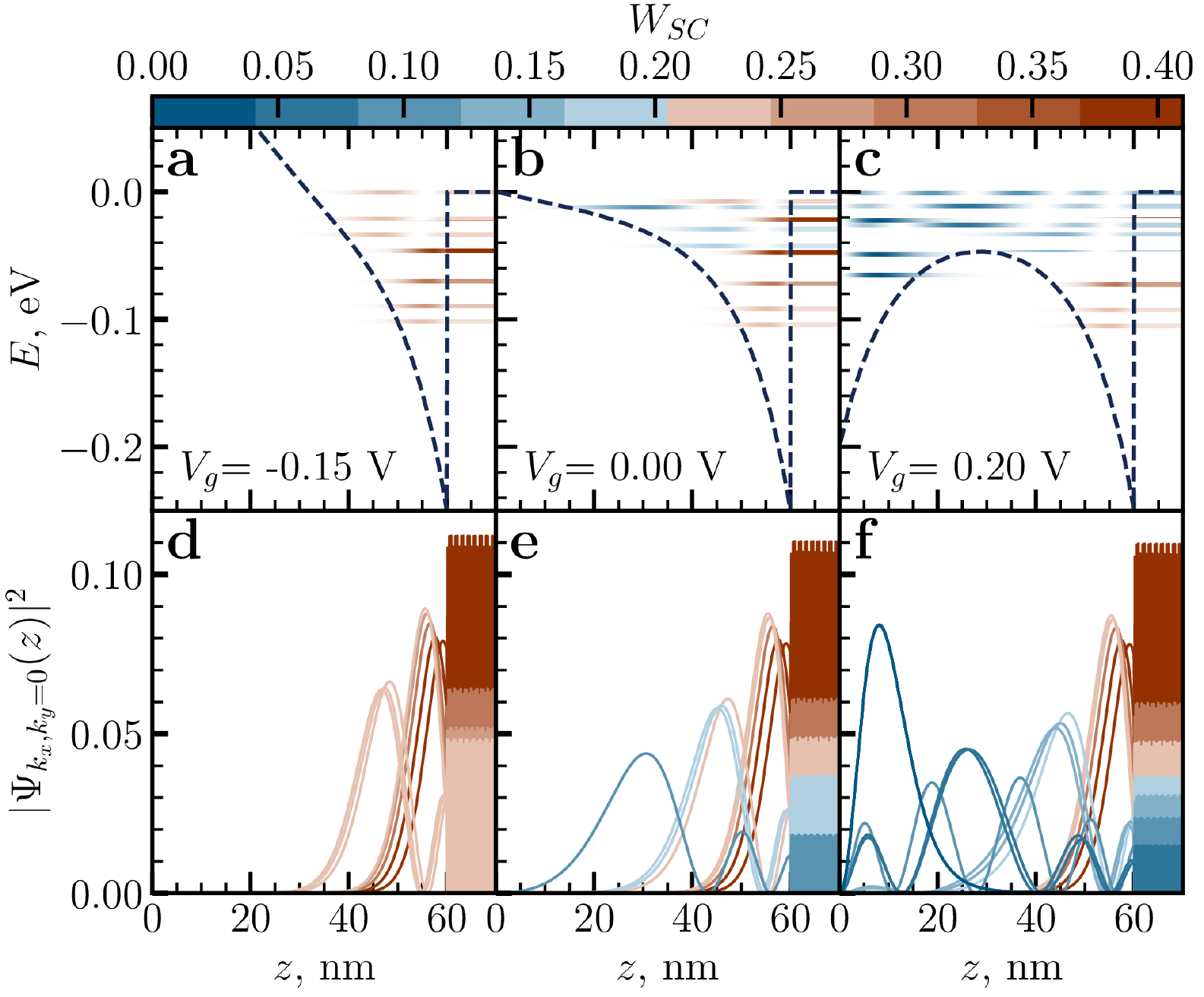}
\caption{(Color online) Eigenstates of the Hamiltonian~(\ref{eq:H_ky}) at $k_x= 0$ for $V_g=-0.15$V, $V_g=0$V, and $V_g=0.2$V. The top panel shows the electrostatic potential for reference purposes, with horizontal lines denoting the bottom of each band below the Fermi energy. The color scale indicates the weight in the superconductor, and in the semiconducting part, the intensity indicates the square modulus of the eigenstate. The lower panel shows the eigenfunctions explicitly with the same color coding. } 
\label{fig:fig5}
\end{figure}

\begin{figure}
\includegraphics[width=0.95\columnwidth]{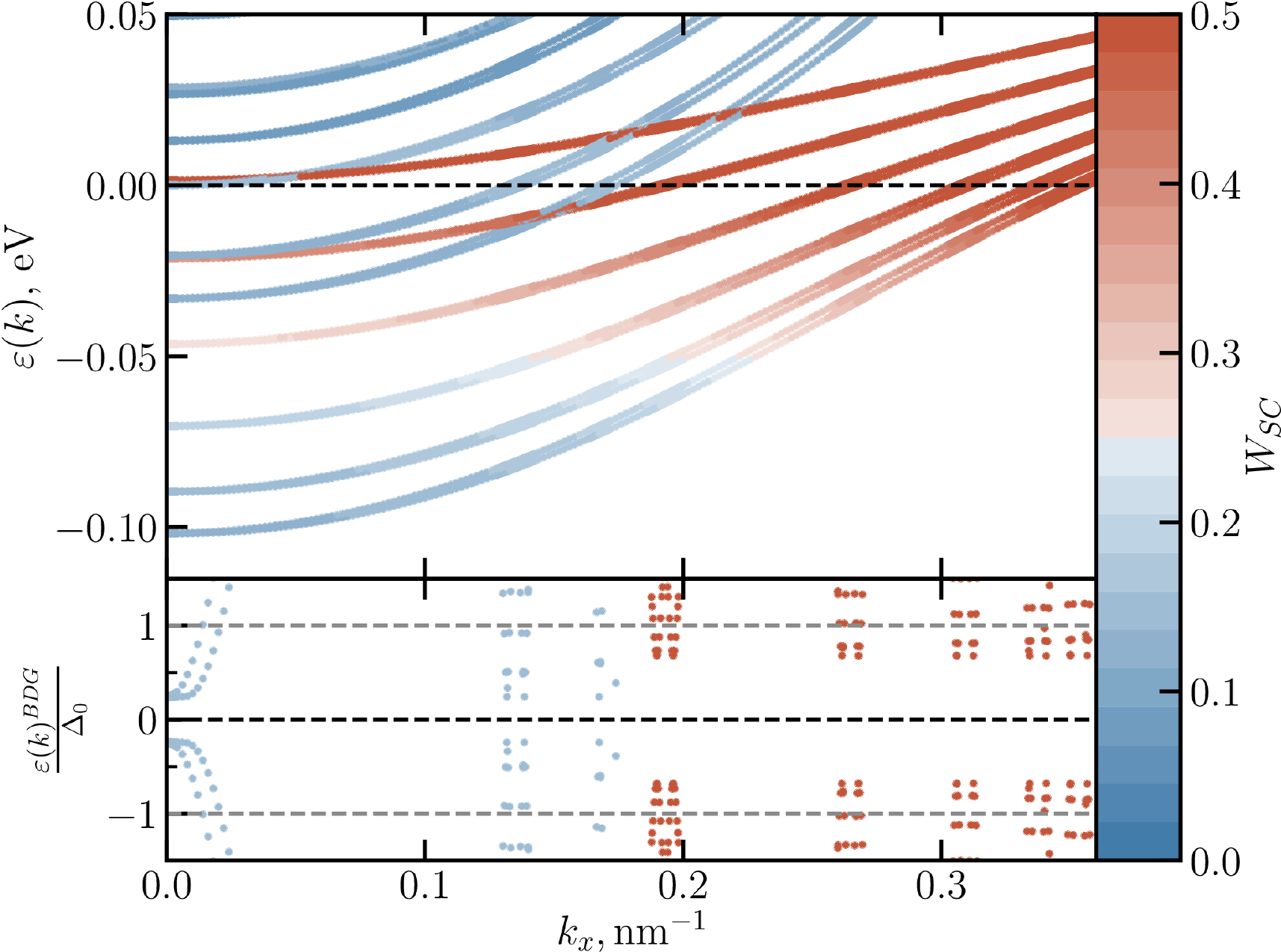}
\caption{(Color online) Top: band structure in the normal state at $V_g = -0.15~\mathrm{V}$. Color indicates the weight of the state in the superconductor. Hybridization between states is seen by the changing color of the subbands. Bottom: band structure of the system in the superconducting state. The induced gap in each subband depends on the hybridization to the superconductor. It can clearly be seen that bands with stronger hybridization (red colors) are characterized by a larger induced gap. Here we used the same parameters as in  Fig.~\ref{fig:fig5}(a) and (d).}
\label{fig:fig6}
\end{figure}

We now discuss the nature of the electronic states in the electrostatic environment determined by the gate as well as the band offset between the semiconducting wire and the metallic shell. In particular,
we will investigate how strongly states are hybridized between the two materials depending on the gate voltage.

In the top panels of Fig.~\ref{fig:fig5}, we show the electrostatic profile (cf. Fig.~\ref{fig:fig4})
for three values of the gate voltage. The lower panels show the square of the wave function  $|\Psi_{k_x=0,n_y}(z)|^2$ for all occupied subbands for the corresponding electrostatic profile.
Here we have chosen the momentum of the band bottom, $k_x=0$, so that all filled bands are included.
The color scale in Fig.~\ref{fig:fig5} indicates the weight of
the wavefunctions in the superconductor (see Eqn.~(\ref{eq:Wsc})).
In the top panels, we have also superimposed horizontal lines showing the energy of the corresponding
subbands; furthermore, the intensity of the lines shows the square magnitude of the wave functions, and
in the semiconducting part the color scale indicates again the weight in the superconductor.

For the case of $V_g=0$ (middle column of panels), we find 9 hybridized subbands, some of which are mostly localized in the SM whereas the others have large weight in the superconductor~\footnote{Other 1D subbands in Al having effective Fermi energies larger than $0.25$~eV are not shown since they do not hybridize with the semiconductor subbands due to energy-momentum conservation. However, when Al is disordered, these subbands  hybridize and, therefore, have to be taken into account. }.

For $V_g<0$ the electrostatic potential confines the wave function in the SM to a very narrow region
close to the SM/Al interface. 
Such confinement favors a strong hybridization of the SM and Al eigenstates, thus giving
rise to states which have large weight in both the SM and Al. Such large hybridization is
prevented in the absence of the confining electrostatic potential due to the 
large mismatch between the Fermi velocities of the two materials. 
The strong confining potential due to
the band offset is therefore critical for the hybridization of the SM and Al states.

For $V_g > 0$, we see in Fig.~\ref{fig:fig5}~(c) that a number of subbands closer to the Fermi
energy appear which are not confined to the interface, and instead have appreciable weight throughout
the SM. These states have very small $W_{sc}$. Their contribution to the density can also be
seen in panel (b) of Fig.~\ref{fig:fig4} in the peak of the density near the gate.

While one might naively expect that the lowest bands are most confined to the interface and thus
hybridize most strongly, this is not reflected in the data shown in Fig.~\ref{fig:fig5}.
To further elucidate which bands most strongly hybridize with the superconductor, we show the
full band structure at $V_g = -0.15~\mathrm{V}$ in the top panel of Fig.~\ref{fig:fig6}. Here, color again
indicates the weight of the state in the SC; however, in contrast to Fig.~\ref{fig:fig5}, we
do not just consider $k_x=0$. We observe that hybridization with the superconductor may depend
strongly on $k_x$, and in this case is generally strongest at large $k_x$.

\subsection{Superconducting properties at $B=0$}

\begin{figure*}
\includegraphics[width=0.95\textwidth]{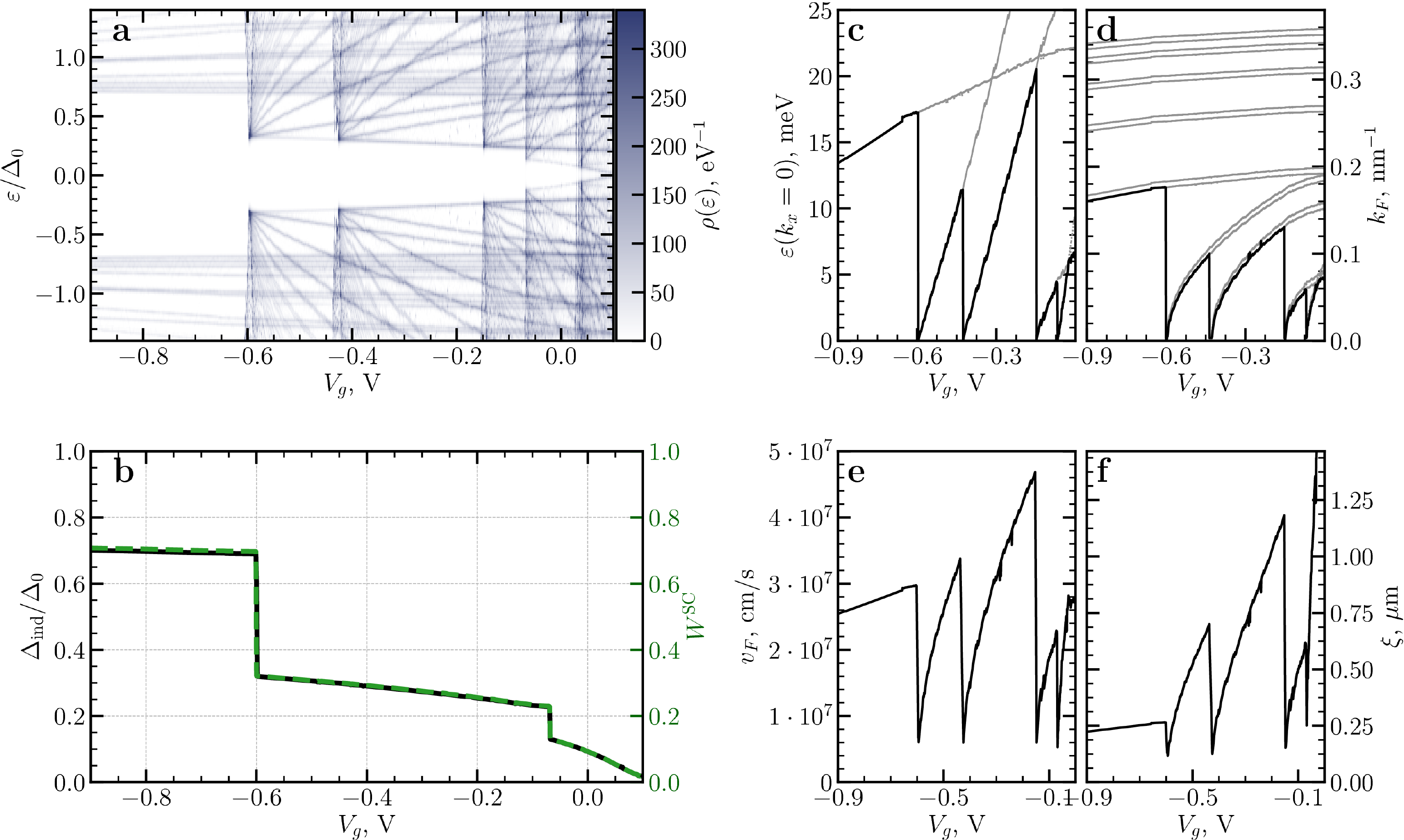}
\caption{(Color online) Characterization of the superconducting state at $B=0$, i.e. in the trivial $s$-wave superconducting phase, as a function of the gate voltage \vg:
(a) Density of states, (b) induced gap (black) and weight in the superconductor (green), (c) Fermi energy, (d) Fermi momentum, (e) Fermi velocity and (f) SC coherence length. The discontinuities correspond to transitions where bands are driven below the chemical potential and thus become occupied (for an illustration, consider the transition between panels (a) and (c) of Fig.~\ref{fig:fig8}). In panel (b), the correspondence between the magnitude of the induced gap and the hybridization between SM and SC (as measured by the $W^{\mathrm{SC}}$) is clearly shown. In panels (c) and (d), all bands are shown in grey, while the occupied band closest to the Fermi energy is highlighted in black. 
}
\label{fig:fig7}
\end{figure*}

We now turn our attention to the situation where the Al shell is in the superconducting state.
The value of $W^{\mathrm{SC}}$ at $k_F^{(n)}$ correlates well with the magnitude of the induced superconducting gap \deltaind for a given subband.
From the discussion of the previous subsection, we can then immediately conclude that different
subbands will have different values of \deltaind.
This is illustrated in the lower panel of Fig.~\ref{fig:fig6}, in which the subbands are shown for the case when $\Delta_0\neq 0$,
for energies of the order of $\Delta_0$.
We see that bands with smaller $W^{\mathrm{SC}}$ (shown as more blue) also have smaller \deltaind.
The smallest value of  \deltaind is what fixes the superconducting gap for the SM-SC hybrid nanowire.
This again emphasizes the importance of the strong confining potential to increase the hybridization
between the two materials and thus a large induced gap.

Figure~\ref{fig:fig7}~(a) shows the evolution of the DOS with $V_g$.
For $V_g<-0.6$~V all subbands in the SM hybridize very strongly with the Al subbands and, thus, have a large induced SC gap. As a result, there are no subgap states below $\varepsilon \approx 0.6 \Delta_0$.
As \vg increases and the electrostatic potential becomes less confining, additional SM subbands become occupied for certain threshold values of $V_g$. As shown in Figure~\ref{fig:fig7}~(a), the number of subbands jumps at $V_g \approx(-0.6,-0.45,-0.15,-0.06,0.03)~\mathrm{V}$.
In some cases the additional subbands have a smaller value of $W^{\mathrm{SC}}$ resulting in a decrease of \deltaind.
From Fig.~\ref{fig:fig7}~(b) we can see that this happens for the \vg threshold values of -0.6 and -0.06~V.
For $V_g>0$, as shown in  Fig.~\ref{fig:fig5}~(d), some of the subbands have states that are not localized
close to the SM-SC interface and for which $W^{\mathrm{SC}}$ is negligible. In this situation \deltaind$\to 0$, and the system becomes gapless. 

The evolution of \deltaind
with \vg is shown in Fig.~\ref{fig:fig7}~(b), together with
the evolution of $W^{\mathrm{SC}}$.
From this figure we see that for \vg$<-0.6$~V, $\Delta_\mathrm{ind} \approx 0.75\Delta_0$.
Furthermore, these results indicate that the evolution of the nanowire's superconducting gap with \vg 
can be quite non-trivial and is closely related to $W^{\mathrm{SC}}$.
In order to have $\Delta_\mathrm{ind} \sim \Delta_0$, strong confining potentials ($V_g<-0.6$~V) are necessary. Conversely,
in the case of a positive gate voltage, there are occupied states in the SM (see right panels of Fig.~\ref{fig:fig5}) which are far away from the SC and, as a result, are weakly proximitized.

Figures~\ref{fig:fig7}~(c)-(f) show the evolution of $\eps_F$, $k_F$, $v_F$, and $\xi$ with \vg for the subband with the smallest induced superconducting gap, which determines \deltaind for the system.
For a fixed number of subbands, as \vg increases $\eps_F$, $k_F$ and $v_F$ grow, see Fig.~\ref{fig:fig7}~(c)-(e).  
Using the values of $v_F$, and \deltaind one can estimate the coherence length 
$\xi= \hbar v_F/(\pi \Delta_{\rm ind})$. 
From Fig.~\ref{fig:fig7}~(b) we see that change in \vg preserving the number of occupied subbands leads to small changes in \deltaind. Thus, 
the variations of $\xi$ are mostly due to the changes in $v_F$, see Fig.~\ref{fig:fig7}~(f). We see that, as long as the number of subbands is constant, $\xi$ grows with \vg and follows $v_F$. The discontinuities in $\eps_F$, $k_F$, $v_F$, and $\xi$ appear when the number of occupied subbands changes, see Fig.~\ref{fig:fig7}~(a)-(f).

\subsection{Superconducting properties at finite magnetic fields}

\begin{figure}
\includegraphics[width=0.95\columnwidth]{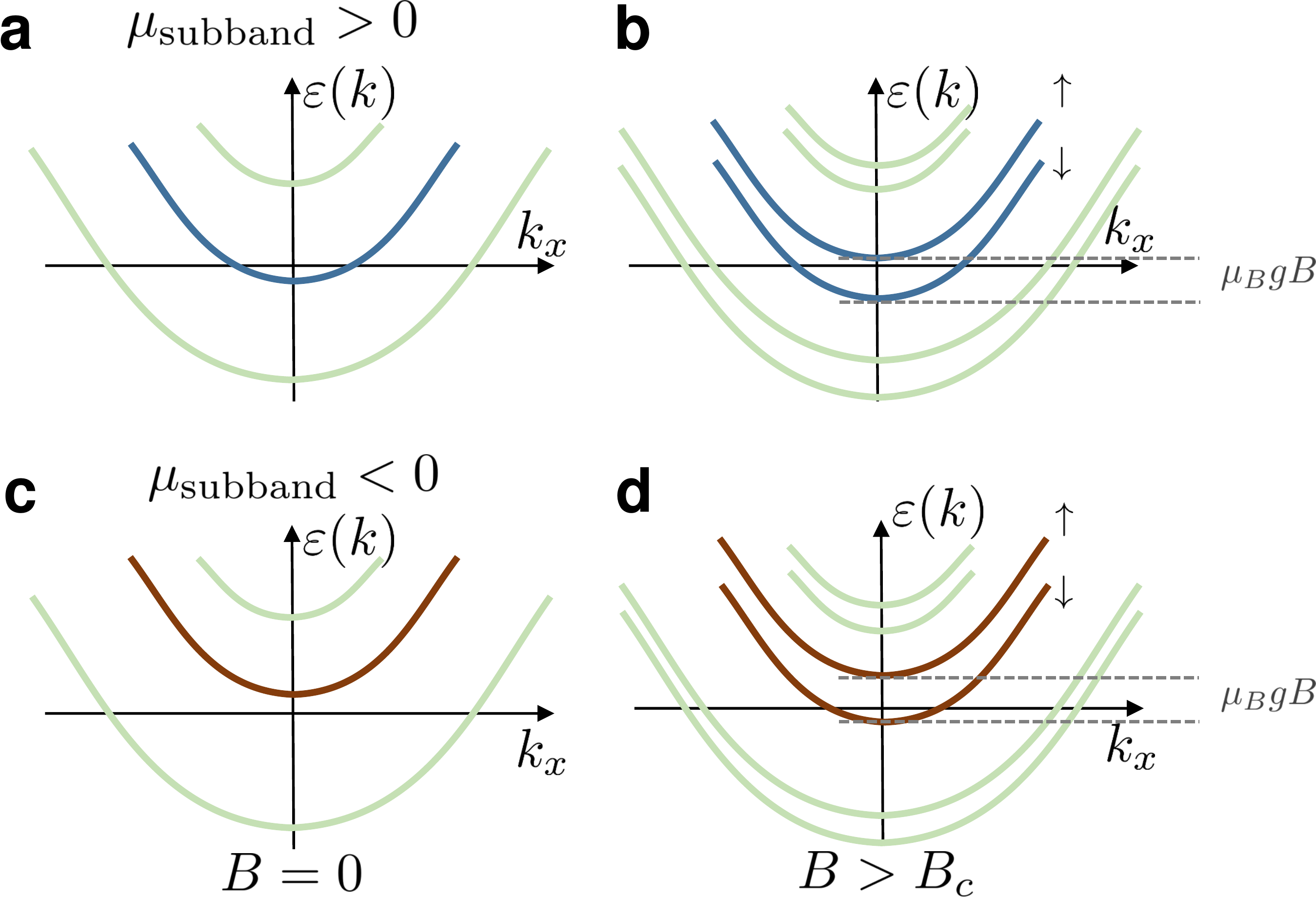}
\caption{(Color online) Illustration of typical band structures, and the two different scenarios for the topological phase transition from an even to an odd number of occupied spin-subbands. The left two panels (a) and (c) show the situation without magnetic field, $B=0$, where a subband is slightly below ((a), one spin-degenerate Fermi point) or above the chemical potential ((c), no Fermi points). Upon turning on a finite field that exceeds the distance from the bottom of these subbands to the chemical potential, the corresponding panel on the right is obtained, where an odd number of spin-split subbands is occupied.}
\label{fig:fig8}
\end{figure}
    
We now study how the properties of the SM-SC nanowire depend on the presence
of the external magnetic field $B$ aligned along the longitudinal direction of the wire.
As discussed in Sec.~\ref{sec:setup}, in our treatment the magnetic field enters only via the Zeeman term. For $B\sim 1~\mathrm{T}$, orbital effect of the applied magnetic field is small since the SC is only 10~nm thick and
in the regime of interest the wave functions in the SM are confined to  the SM-SC interface within 20~nm range. 

We start by investigating Zeeman splitting for the nanowire with multi-subband occupancy. The corresponding band structure is shown in Fig.~\ref{fig:fig8}~(a). Let's consider the gate voltage such that the highest occupied subband (shown in blue color) has small Fermi energy. The application of a magnetic field splits the subband and, at some critical field $B_c$, drives the minority subband across the Fermi level (provided $B_c$ is less than the critical field of the superconductor). This is illustrated in Fig.~\ref{fig:fig8}~(b). At this point, the majority subband  becomes the highest occupied band, and, thus, many properties such as the Fermi energy, Fermi velocity and Fermi momentum change discontinuously.

Another scenario corresponds to Fig.~\ref{fig:fig8}~(c), where a band is just above the chemical potential. In this case, the gap of the system is determined by the lower occupied subband (shown in green). An increasing magnetic field splits the lowest \emph{un}occupied band (shown in red) and eventually it becomes occupied. 

\begin{figure}
\includegraphics[width=0.95\columnwidth]{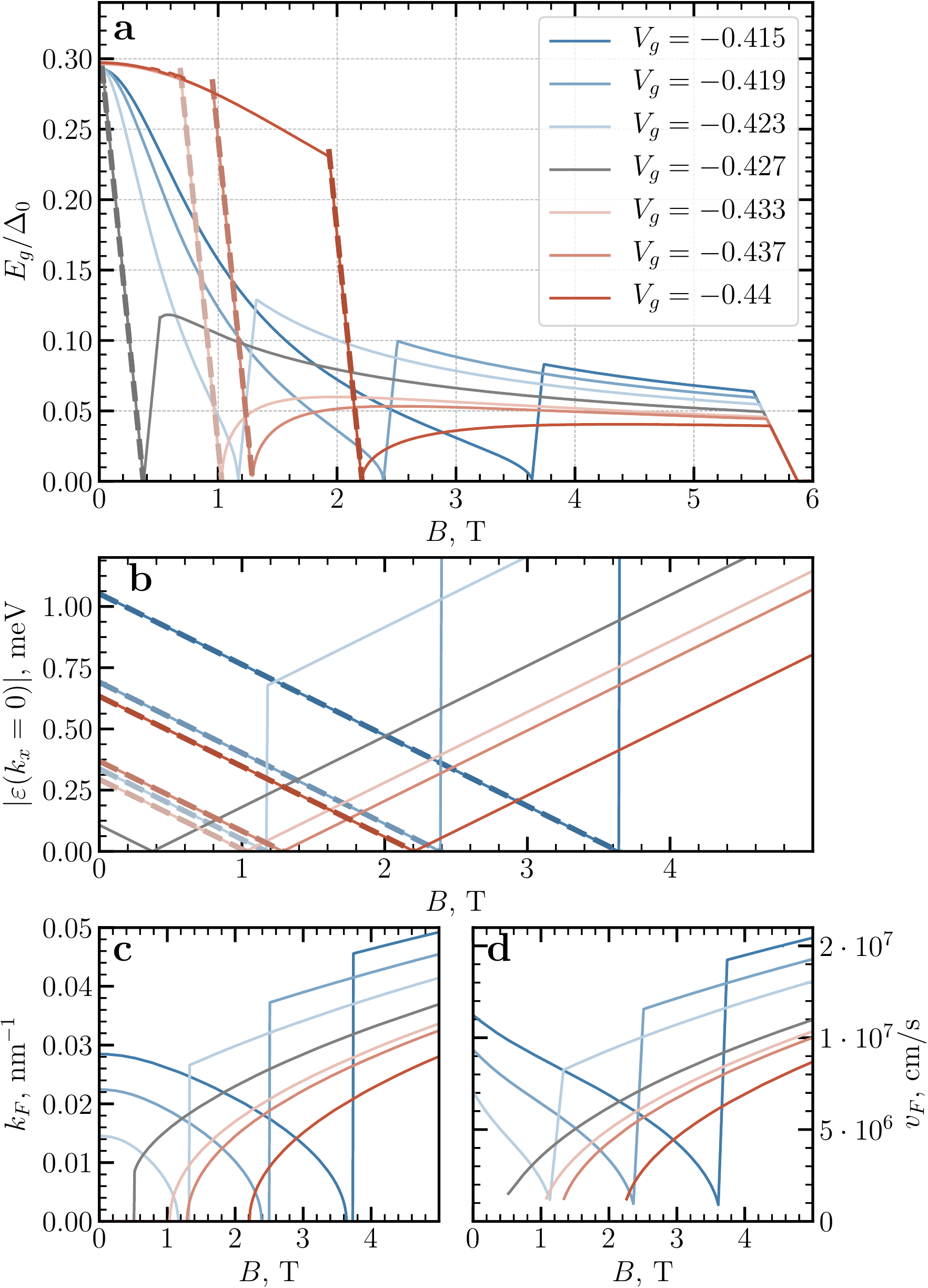}
\caption{ (Color online) The evolution of the system parameters as a function of a magnetic field $B$. Estimates for the (a) gap, (b) absolute value of the effective Fermi level of the subband $|\varepsilon(k_x = 0)|$ (where $k_x=0$ is the location of the band bottom), (c) Fermi momentum and (d) Fermi velocity as a function of magnetic field. Colors correspond to different gate voltages: for blue lines, the gate voltage is more positive, leading to a situation as sketched in the upper two panels of Fig.~\ref{fig:fig8}. Red lines, on the other hand, have a more strongly negative gate voltage, thus leading to the situation of the two lower panels in Fig.~\ref{fig:fig8}. Gray color corresponds to the threshold value $V_g = -0.427$~V, at which the subband is characterized by a vanishing effective chemical potential.}
\label{fig:fig9}
\end{figure}

In both these cases, we end up with an odd number of occupied subbands at large enough magnetic fields and, thus, the nanowire can be driven into the topological regime. However, the evolution of the gap with the magnetic field is drastically different in these two cases. This can seen in Fig.~\ref{fig:fig9} which shows the evolution with magnetic field of the spectral gap $E_g$, effective Fermi energy for the highest occupied subband, and corresponding $k_F$, $v_F$, and $\xi$ for different gate voltages close to the the threshold value $V_{g,t}=-0.427$~V. This threshold value, corresponding to the gray curves in Fig.~\ref{fig:fig9}, corresponds to the gate voltage at which the relevant subband has exactly zero effective Fermi energy. As $B$ increases from zero the gap $E_g$ decreases and eventually vanishes at $B=B_c$ corresponding to the topological phase transition. For $B> B_c$, the nanowire is driven into the spinless regime with p-wave pairing potential. The p-wave gap exhibits a non-monotonic dependence on the magnetic field, and eventually vanishes because s-wave superconductor becomes normal. For our parameters this occurs at $B_{SC}=5.8$~T. Note that we do not take into account orbital effects here  so in practice $\Delta_0$ may vanish before that.        

The blue family of curves in Fig.~\ref{fig:fig9}~(a) correspond to the case when a highest-occupied subband has a small Fermi energy at zero field (top panels of Fig.~\ref{fig:fig8}). In that case,
the gap at zero field is already set by the band that will eventually be split to give rise to topological superconductivity, and, thus, the gap evolves as a smooth function for $B < B_c$. At $B = B_c$ the minority subband crosses the Fermi level, and the topological gap is opened in the majority subband. As dicussed above, the properties of the Fermi points evolve discontinuously across the transition (panels b,c,d), and the gap increases rapidly into the topological phase (panel a).

At more negative gate voltages, the situation shown in Fig.~\ref{fig:fig8}~(c) and (d) is realized, corresponding to the red lines in Fig.~\ref{fig:fig9}. Here, the gap at $B=0$ is determined by the next occupied subband. Upon the application of a magnetic field, the distance between the majority subband and the chemical potential eventually becomes smaller than the gap induced in the next-highest subband. This distance thus sets the spectral gap. The discontinuity of the gap function can be seen in Fig.~\ref{fig:fig9}~(a). At $B>B_c$ the topological gap is opened in the majority subband. In this cases, we plot in Fig.~\ref{fig:fig9}  the properties of the Fermi points only for the subbands that eventually become topological, and thus plot no values below the topological phase transition.
    
It is very interesting to notice that the size of the induced superconducting gap for $B=0$ 
does not necessarily correlate with the size of the topological gap. 
This can be understood from the fact that the topological gap for $B>B_c$ is always opened in the same band, whereas the \deltaind at $B=0$ is opened in a different band when \vg becomes smaller than $V_g=-0.427$~V. 
As can be seen from Fig.~\ref{fig:fig9}~(c) the Fermi momentum $k_F$ for $B>B_c$, which corresponds always to the same band, increases with \vg. The topological gap increases with $k_F$ since the effective Rashba field is stronger at higher momentum, allowing the $s$-wave pairing to induce a larger gap.
This is a very important result because \deltaind at $B\gtrsim B_c$ sets a crucial scale for the robustness
of a topological qubit against error sources such as thermal fluctuations, diabatic corrections, disorder~\cite{Meng2012, liu2017,Knapp2017, Fu2018} etc.

\subsection{Effective g-factor}

\begin{figure}[ht]
\includegraphics[width=0.95\columnwidth]{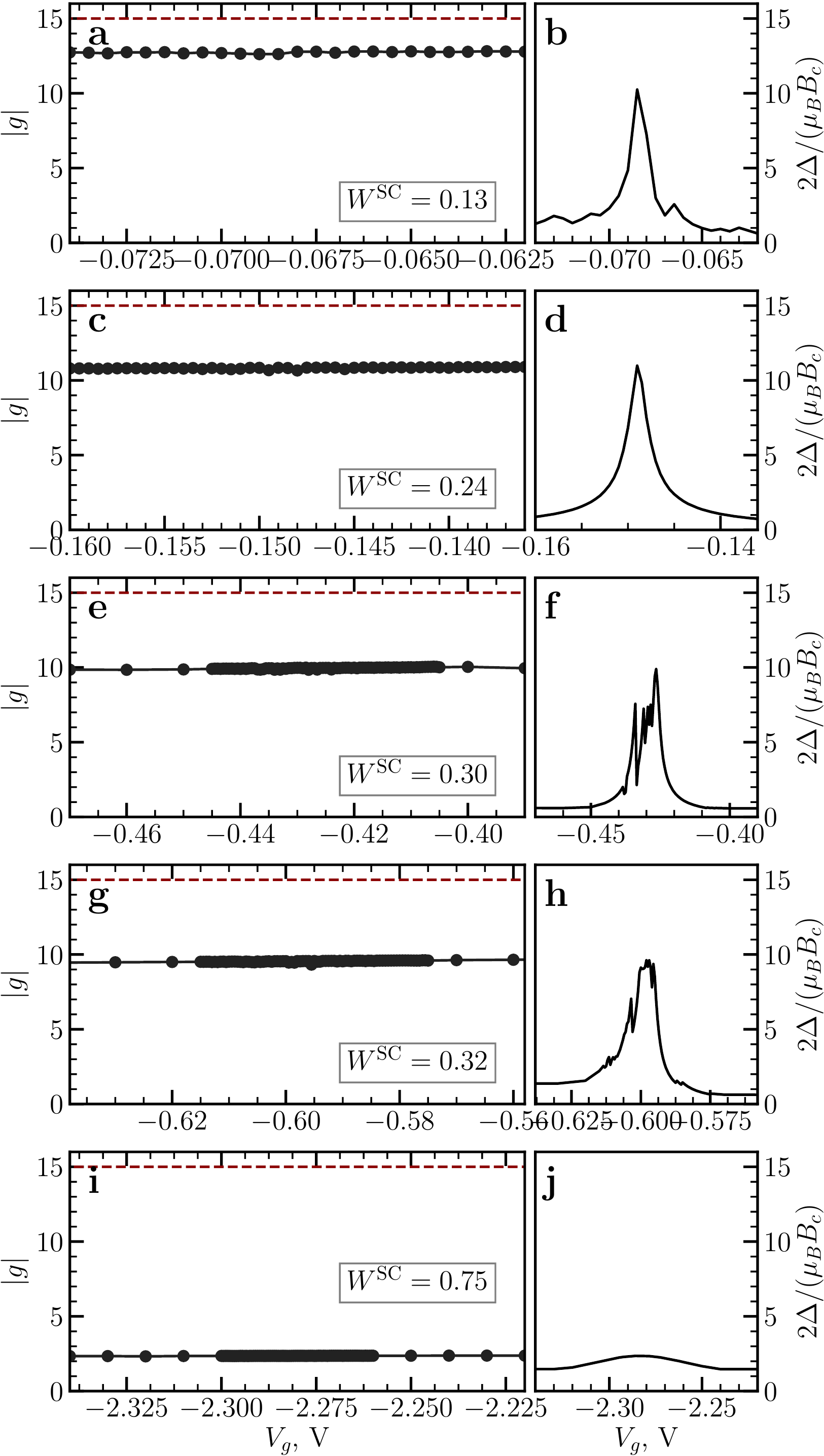}
\caption
{(Color online) 
Absolute value of g-factor (left column) and ratio of the induced gap to the critical magnetic field $2\Delta/(\mu_B B_c)$ (right column) as a function of gate voltage in the vicinity of threshold gate voltage values $V_g = -0.07$ (a,b), $-0.15$ (c,d), $-0.45$ (e,f), $-0.6$ (g,h), $-2.3$ (i,j) at which the number of subbands change in the system. The red dashed line is the absolute value of bare semiconductor g-factor $|\gbare_{SM}|$. 
}
\label{fig:fig10}
\end{figure}

A crucial quantity to characterize the semiconductor-superconductor system is the effective g-factor of
the hybrid system. Due to the drastically different g-factors in the two materials, this will depend intricately
on the wave function hybridization between them. Furthermore, the g-factor is crucial in enabling a large and robust topological
phase, since a large g-factor is necessary for the topological phase transition to occur at a magnetic field well below
the critical value at which the Al shell is driven normal. A large effective \gf is thus very helpful in achieving
a sufficient separation between these scales.

The effective g-factor can be obtained from studying the Zeeman splitting of bands at $k_x = 0$, as illustrated in Fig.~\ref{fig:fig8}. In particular, since at $k_x=0$ the spin-orbit terms in the Hamiltonian~(\ref{eq:H_ky}) vanish, the spin-splitting of the bands at $k_x=0$ is entirely determined by the Zeeman term. As the change of the energy levels $\varepsilon(k_x = 0)$ is linear with the magnetic field, the absolute value of the g-factor can be extracted as $|g| = 2\frac{d\varepsilon(k_x=0)}{d\mu_B B}$. This linear fit for $B < B_c$ is illustrated in Fig.~\ref{fig:fig9}b with dashed lines. Note that when the gate voltage is such that the closest subband to the Fermi level is unoccupied (case (c) in Fig.~\ref{fig:fig8}), the slope near the gap-closing is the same as the one at $\varepsilon(k_x=0)$, allowing for a reliable extraction of the g-factor from the tunneling conductance measurements ~\cite{Sole2017}. This is shown in Fig.~\ref{fig:fig9}a by the dashed lines.  

In Fig.~\ref{fig:fig10} we study the dependence of the extracted g-factor on the applied gate voltage in the vicinity of the threshold values at which the number of subbands change (see  Fig.~\ref{fig:fig7}). As expected, we find every subband to be characterized by an almost constant g-factor, with significant changes occuring only at transitions between bands. When the hybridization between semi- and superconductor is weak, the g-factor is close to the bare semiconductor value $|\gbare_{SM}|=15$. Conversely, when the voltage is very negative (the value from Fig.~\ref{fig:fig7} is written in every panel) and the hybridization between semi- and superconductor is strong the g-factor is almost as small as the bare superconducting g-factor $|\gbare_{SM}|=2$. 

Additional information can be extracted from the ratio of the induced gap to the critical field, shown in the right column of Fig.~\ref{fig:fig10}. This quantity is easily accessible in experiments, and has been used in the experimental literature as a proxy for the g-factor~\cite{Sole2017}. Our results clearly show that unlike the g-factor, this quantity has a strong dependence on gate voltage over relatively small gate voltage variations. In particular, a resonant structure appears with a peak that corresponds to the gate voltage being tuned to the threshold value at which the subband crosses the effective chemical potential. Only at this point does this quantity reaches the values of the effective g-factor shown in the left column. 

\begin{figure}
\includegraphics[width=0.95\columnwidth]{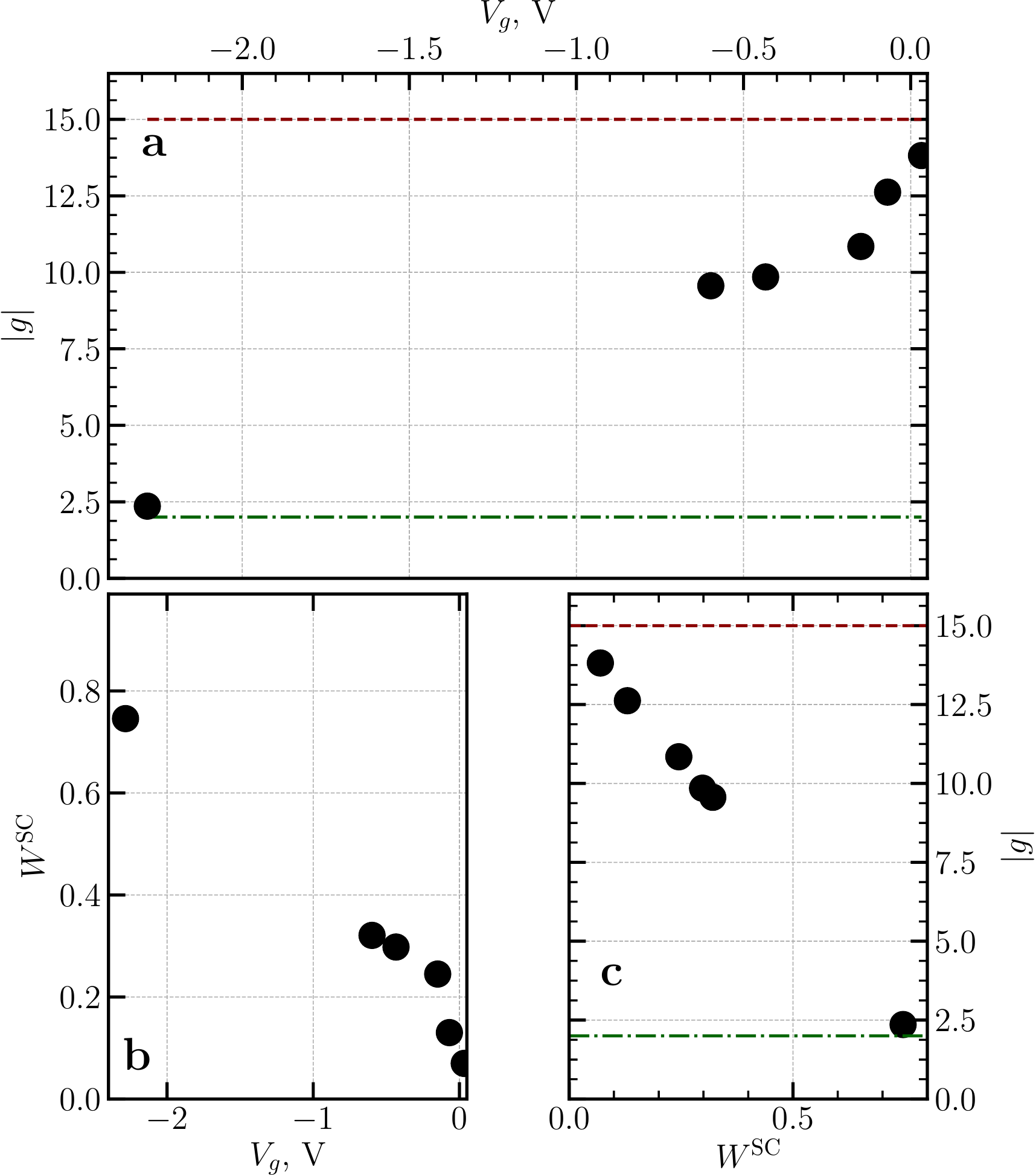}
\caption{(Color online) 
Gate voltage dependence of (a) the absolute value of effective g-factor and (b) the weight in superconductor. Panel (c) relates $|g|$ to the weight in the superconductor $W^{\mathrm{SC}}$. Dots represent typical values for the highest occupied band at a given gate voltage. The dashed horizontal values are the absolute values of the bare g-factor in the semiconductor (red) and superconductor (green). When the coupling to the superconductor is weak, as indicated by a small weight of the wavefunctions in the SC, the g-factor is close to the bare InAs value. The opposite limit occurs in the strongly hybridized regime at more negative gate voltages.
} 
\label{fig:fig11}
\end{figure}
    
Figure~\ref{fig:fig11}~(a) shows the value of \gf for different topological regions (see
also the discussion in Sec.~\ref{sec:pd}). We see that 
as \vg becomes more negative the g-factor becomes smaller and approaches the value of \gf in the SC.
As stated above this is due to the fact that as \vg becomes more negative
the hybridization between SM and SC states becomes stronger as clearly
shown by the evolution of $W^{\mathrm{SC}}$, see Fig.~\ref{fig:fig11}~(b). 
Larger negative values of \vg create an electrostatic potential that more strongly
confines the SM states at the SM-SC interface. The tighter confinement
results in a stronger hybridization between SM and SC states. 
Figure~\ref{fig:fig11}~(c) summarizes the important relation between strength
of the hybridization between SM and SC states and the g-factor by showing the 
dependence of $|g|$ on $W^{\mathrm{SC}}$. 
We see that qualitatively \gf scales linearly with $W^{\mathrm{SC}}$.

\subsection{Topological phase diagram} \label{sec:pd}

\begin{figure}
\includegraphics[width=0.95\columnwidth]{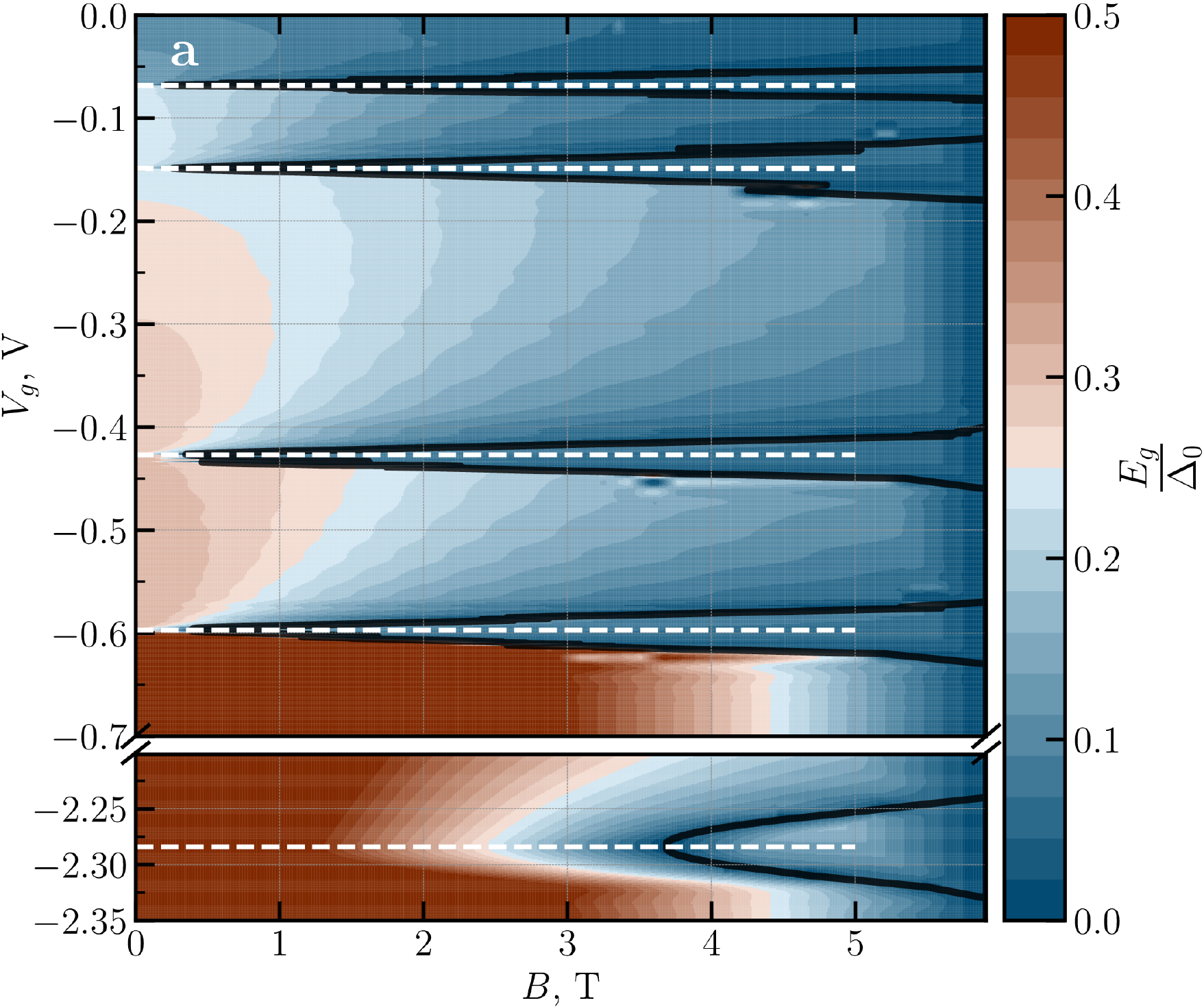} \\
\hspace{-0.5in} \includegraphics[width=0.8\columnwidth]{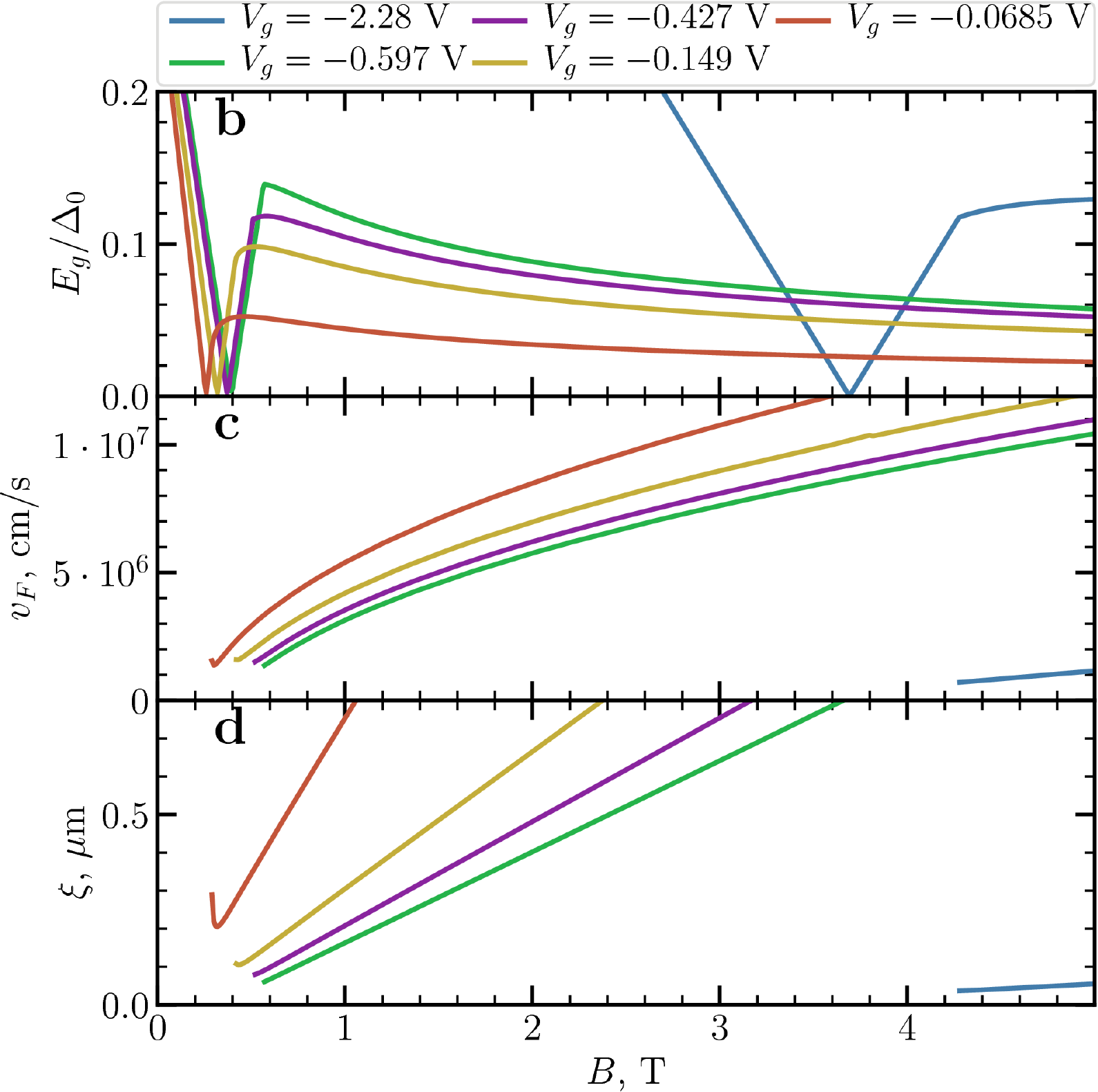}
\caption{(Color online) (a) Topological phase diagram and magnitude of the spectral gap over a range of gate voltages from zero to very strongly negative. Clearly, the gap at zero field is largest for negative gate voltages where hybridization with the Al shell is strongest. The boundaries of the topological phase are marked with solid black lines. Dependence of the (b) gap normalized to the bare Al gap $E_g / \Delta_0$ and estimates for (c) Fermi velocity and (d) coherence length as a function of magnetic field for the gate voltages indicated by dashed lines in (a). }
\label{fig:pd}
\end{figure}

Figure~\ref{fig:pd}~(a) shows the topological phase diagram in the $(V_g,B)$ plane. 
This is one of the most important results of this work: it relates the nature of the superconducting state of the quasi-1D hybrid nanowire to the experimentally relevant and tunable quantities -- the gate voltage and the external magnetic field --
rather than more abstract quantities such as the Fermi energy of the subbands and the Zeeman splitting, which are dependent on applied electric field.
As discussed above, the relation between \vg and the parameters characterizing the nanowire band structure, such as the subbands' Fermi energy and induced superconducting gap,
is highly non-trivial given the nonlinear nature of the \ps problem and the presence of multiple subbands.
For this reason, simplified models in which the subband chemical potential is assumed to be directly
proportional to \vg in general cannot be used to obtain a reliable phase diagram in the 
$(V_g,B)$ plane.
Similarly, we have shown that for the g-factor we cannot take the bare value of the SM.
One qualitative feature that emerges from the results shown in 
Fig.~\ref{fig:pd}~(a) is how the shape and size of the topological regions depend on \vg.
We see that for very large negative \vg the critical magnetic field is higher than that
for small negative values of \vg. This is due to the fact that the hybridization 
of the SM's and SC's states is stronger for larger negative \vg and therefore
the effective magnitude of \gf is much smaller than $g^{\rm bare}_{SM}$ 
causing an increase of the critical field.

The topological phase diagram can be obtained by calculating the topological index $M \in \mathbb{Z}_2$ (Majorana number)~\cite{Kitaev01,Lutchyn2011,Stanescu2011}: 
\begin{align}
M= \sgn\left[\Pf \left( B\left(k_x=0\right) \right)\right] \sgn\left[\Pf \left(B\left(k_x=\pi/a\right)\right)\right], 
\end{align}
where $B(k_x)$ is an antisymmetric matrix which defines the Hamiltonian
of the system in the Majorana basis. The negative/positive sign of $M$ corresponds to a topologically trivial/non-trivial phase. The latter supports Majorana zero modes at the ends of the nanowire. In the continuum limit,  the lattice spacing $a \rightarrow 0$, and the sign of $\Pf \left(B(k_x\rightarrow \infty)\right)$ is fixed. Thus, the topological quantum phase transition corresponds to a change of $\sgn [\Pf \left( B(k_x=0) \right)]$. Note that the topological phase transition in this case is accompanied by a vanishing quasiparticle gap at $k_x=0$ which provides another way of determining of the phase boundary.  

It is illuminating to compare the phase diagram of Fig.\ref{fig:pd}~(a) with previous studies~\cite{Lutchyn2011,Stanescu2011}. Adapting the results of Refs.~\onlinecite{Lutchyn10, Oreg10, Lutchyn2011}, the critical magnetic field for the topological transition is given by 
\begin{equation}
 B_c=\frac{\sqrt{[\eps(k_x=0)]^2+\Delta^2}}{|g|},
 \label{eq:pd2}
\end{equation} 
where $\eps(k_x=0)$ defines the position of the band bottom at $k_x=0$ relative to the Fermi energy in the superconductor and $g$ is the effective g-factor. As follows from the discussion above, the effective g-factor can be obtained by calculating Zeeman splitting at $k_x = 0$.

\begin{figure}
\includegraphics[width=0.95\columnwidth]{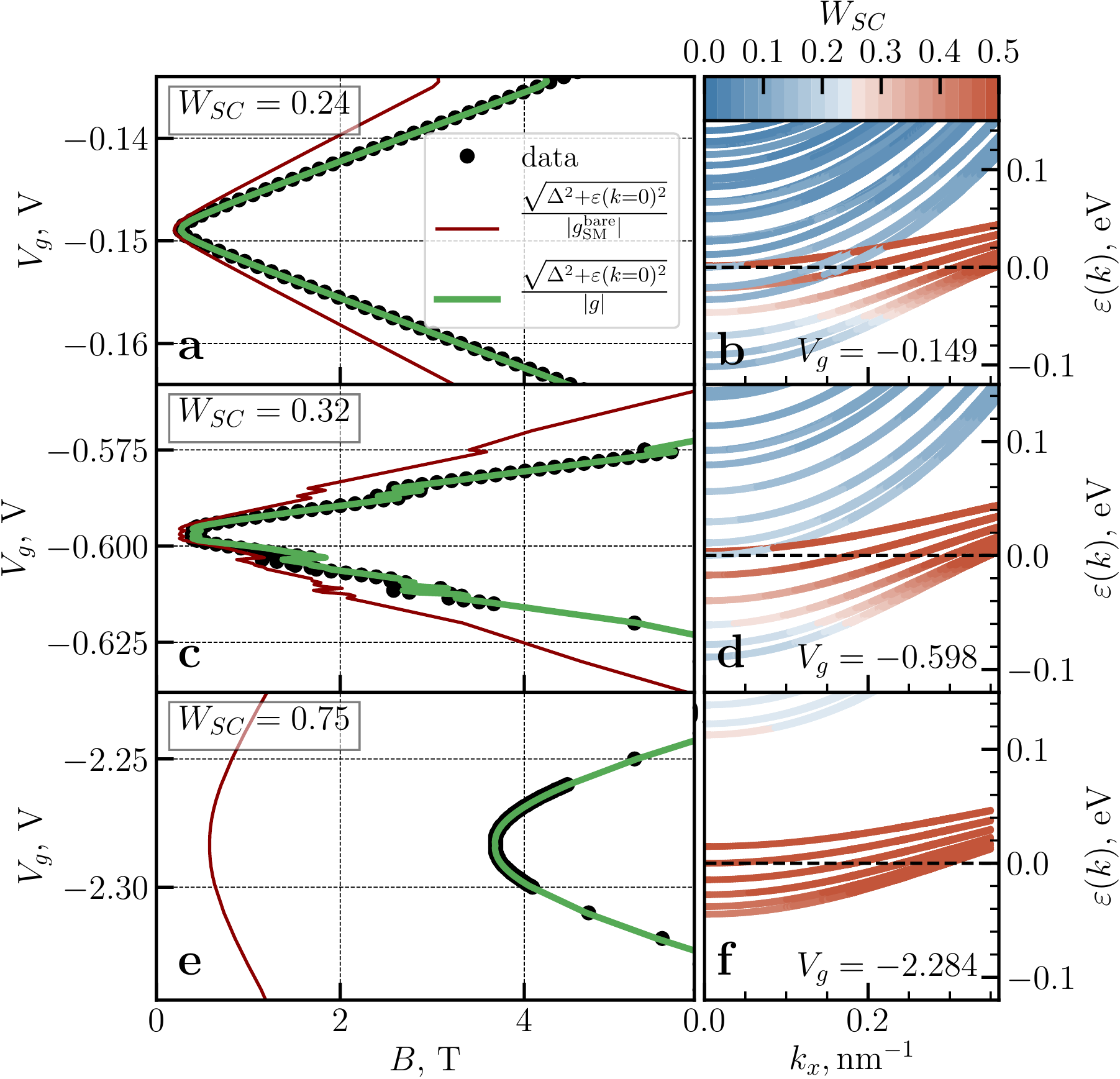}
\caption
 {
  Phase diagram, panels~(a,c,e), and band structure, panels~(b,d,f) in the vicinity of threshold gate voltages, corresponding to the change in the number of subbands: $V_g = -0.15$ (a,b); $V_g = -0.6$ (c,d); $V_g = -2.3$ (e,f). Black dots in the left panels correspond to the numerically calculated phase boundary. Red and green curves in left panels are the estimates obtained using the standard relation~\ceq{eq:pd2} and the bare SM's g-factor for the red curves and the renormalized g-factor Fig.~\ref{fig:fig10}, for the green ones. 
}
\label{fig:fig13}
\end{figure}

Having obtained the dependence of $\eps(k_x=0)$, $g$ and $\Delta$ on \vg, 
one can draw the boundaries in the $(V_g,B)$ plane of the topological phase diagram using Eq.~\ceq{eq:pd2}. These boundaries are shown in green in the left panels of Fig.~\ref{fig:fig13}. We see that they exactly match the boundaries obtained by identifying the value of $B$, $B_c$, where the gap is closing $\Delta_{\rm ind}=0$. In particular, Eq.~\ceq{eq:pd2}
gives the correct boundaries {\em if} the renormalization of the g-factor is taken into account. 
On the other hand, if in Eq.~\ceq{eq:pd2} we use the bare SM g-factor,
Eq.~\ceq{eq:pd2} gives incorrect boundaries,
shown in red in the left panels of Fig.~\ref{fig:fig13}.
The boundaries obtained using the bare SM g-factor
overestimate the size of the topological region, especially when \vg is strongly
negative, as shown in the bottom left panel of Fig.~\ref{fig:fig13}.
As discussed above, this is due to the fact that the value of the g-factor, in the strong coupling regime,
is strongly renormalized by the hybridization between the SM and SC states.
The right panels in Fig.~\ref{fig:fig13} show the band structure of the SM-SC nanowire close to the Fermi energy when $\Delta_0\to 0$
for the appropriate values of \vg.
We see that for very negative values of \vg, bottom panel, the SM states are very strongly hybridized with the SC states. This
result is consistent with the fact that for this case \gf is much smaller than $g_{SM}$ and therefore
the topological region is much smaller that we would have obtained assuming $g=g_{SM}$.

Figures~\ref{fig:pd}~(b)-(d) show the dependence of the gap (b), Fermi velocity (c), and coherence length (d) on the magnetic field, for $B>B_c$, 
for different domes in the topological phase diagram shown in Fig.~\ref{fig:pd}~(a). To obtain these figures we use the representative gate voltages
indicated by the white dashed lines in Fig.~\ref{fig:pd}~(a).  
For fixed magnetic field, the gap decreases with the gate voltage while the Fermi velocity increases. As a consequence, the coherence length increases.
The results of Fig.~\ref{fig:pd}~(d) show that when the topological gap is maximal, $\xi$ has values in the $100-300$~nm range, and
that $\xi$ grows linearly with $B$ for $B>B_c$.
The growth of $\xi$ with $B$ is slower for more negative gate voltages, a reflection of the fact that for
larger negative gate voltages the effective g-factor is smaller due to the stronger hybridization of the SM's and SC's states.
The results of Fig.~\ref{fig:pd}~(d) are important for the design of Majorana-based qubits since their topological protection relies on the exponentially-small splitting of Majorana zero modes which, in turn, strongly depends on the coherence length.

\subsection{Effect of disorder in SC-SM heterostructures} \label{sec:disorder}
\begin{figure}[!ht]
\includegraphics[width=0.95\columnwidth]{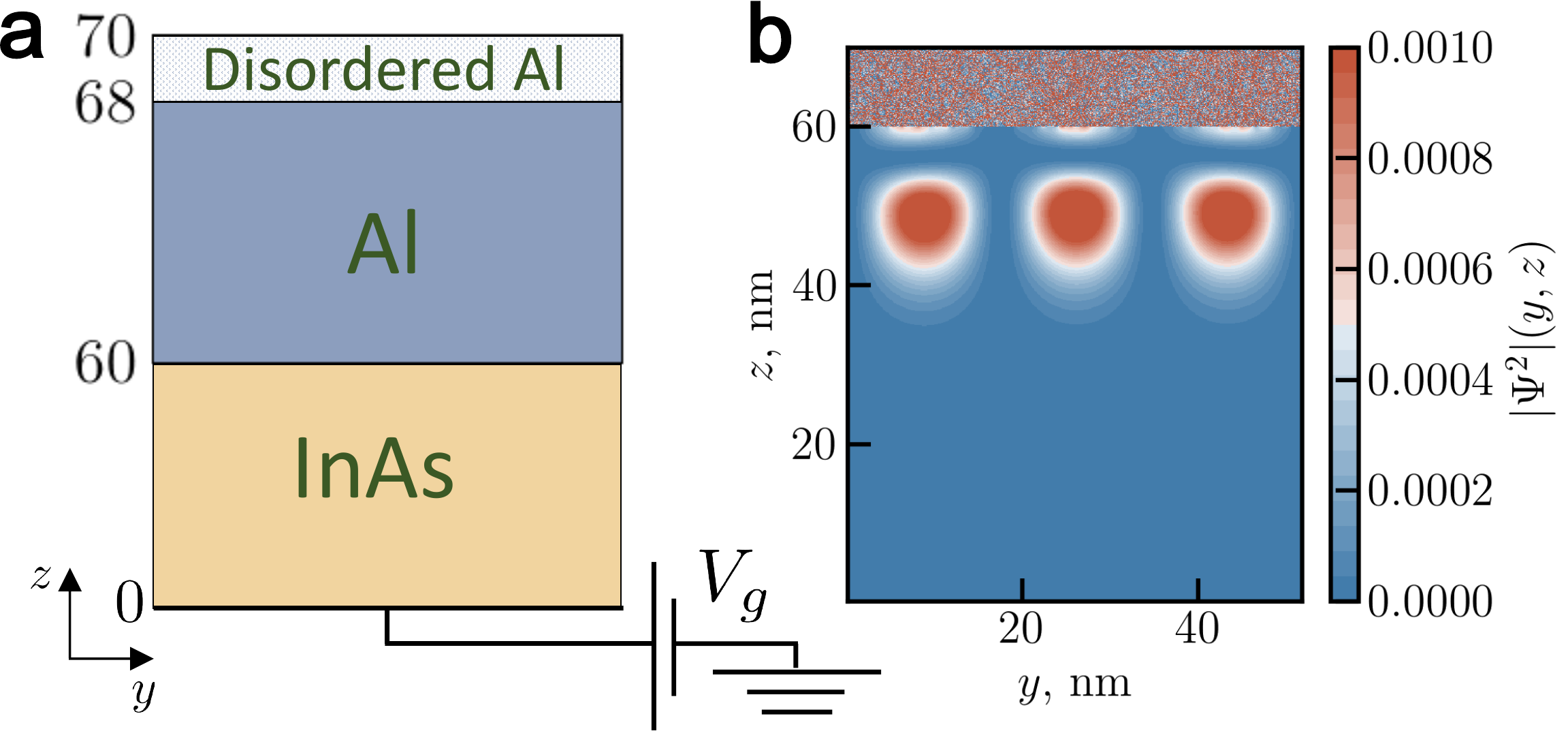}
\caption
 {
    (Color online) (a) SM-SC heterostructure in the slab geometry in the presence of disorder on the surface of superconductor (not to scale); (b) Square modulus of the eigenstate in the presence of disorder highlighting the enhanced scattering in the superconductor. 
}
\label{fig:sketch_dis}
\end{figure}

Disorder is ubiquitous in solid-state systems and has a strong impact on the physical properties of proximitized nanowires. It has been shown that disorder in the SM and at the SM-SC interface is detrimental to the topological phase~\cite{Potter2011, Potter2011a, Lobos12, Lutchyn2012, Sau2012, Tkachov2013, Sau2013, Hui2015, Cole2016, DongLiu2017}. However, in the MBE-grown SC-SM heterostructures~\cite{Krogstrup2015,Lutchyn2018} disorder effects are minimized resulting in high quality SMs as well as SM-SC interfaces. The SC (i.e. Al) is also nominally of high quality but its outer surface is covered by an amorphous oxide layer, see Fig. \ref{fig:sketch_dis}. Therefore, the scattering from the outer boundary randomizes the motion of quasiparticles in the SC. Due to the large effective mass mismatch and the conservation of the momentum at SM-SC interface, disorder in the SC is not expected to interfere with the observation and manipulation of MZMs~\cite{Lutchyn2012, DongLiu2017, Sticlet2017}. Nevertheless, as we will show it can strongly affect key quantities 
of the proximitized nanowire, such as \deltaind and the critical field $B_c$, as well as their dependence on the external gate voltage. 

We model disorder by adding the random potential $V_D$ to the Hamiltonian~\ceq{eq:H_n}. Due to the computational complexity of the problem we consider here the disorder potential $V_D (z,y)$ (i.e. homogeneous along the wire). Such potential hybridizes different $y-$ and $z-$subbands which effectively increases density of states in the SC. In contrast to Eq.~(\ref{eq:H_ky}) the modes in the $y-$direction are not separable anymore, and we have to resort to a numerical solution of the full Hamiltonian, leading to a much greater numerical complexity. We calculate results for a given disorder realization and then average physical observables over approximately 25 disorder realizations. 

We model the amorphous oxide layer in the superconductor by adding a disorder potential within $l_z=2$~nm from the outer surface of the SC, see Fig. \ref{fig:sketch_dis}. We assume the disorder potential to have zero average and to be spatially uncorrelated: 
\begin{align}
 \langle \langle V_D \rangle \rangle&=0 & \langle \langle V_D(\rr)V_D(\rr') \rangle \rangle &= K_0 n_{\rm imp}\delta(\rr-\rr')
 \label{eq:VD02}
\end{align}
Here $\langle \langle...\rangle \rangle $ denotes averaging over disorder realizations, $K_0$ parametrizes 
the disorder strength, and $n_{\rm imp}$ corresponds to the density of impurities. Considering the finite spatial resolution in the numerical calculation and the uniform box distribution of the disorder with the amplitude  $U_{\rm D}$, $K_0$ is related to $U_{\rm D}$ as $\delta\rr U_D^2/3=K_0 n_{\rm imp}$, where
${\delta\rr}$ is the volume of a single cell of the spatial discretization (see Table \ref{table:params}), that we set to be uniform in the region where the disorder is located. We vary $U_{\rm D}$ in the parameter range $0-1$~eV which corresponds to the effective mean-free length larger than 10nm. 

\begin{figure}[t]
\includegraphics[width=0.95\columnwidth]{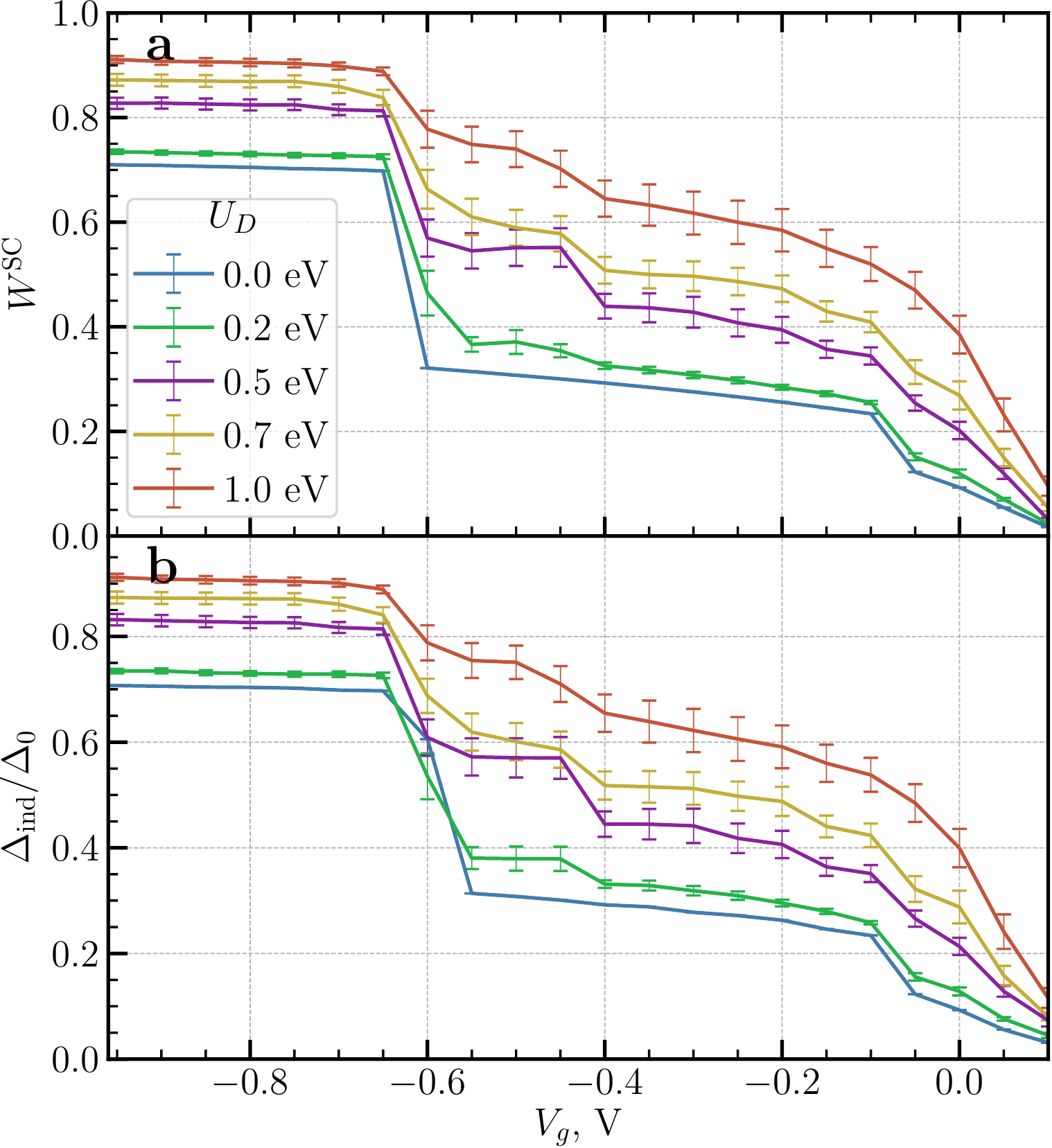}
\caption
 { (Color online) (a) Weight in the superconductor and (b) induced gap as a function of the gate voltage for a range of disorder strengths $U_D = 0-1~\text{eV}$ averaged over 25 disorder realizations. The presence of disorder increases the induced gap. 
}
\label{fig:delta_dis}
\end{figure}

One of the main effects of the disorder in the SC is to break the conservation of the momentum and to induce broadening of the SC subbands which effectively increases the number of superconducting subbands hybridizing with a given SM mode. This effect is shown in Fig.~\ref{fig:sketch_dis}b where we plot the wavefunction probability density: the wavefunction probability in the SC is random, which corresponds to chaotic motion of quasiparticles, whereas the one in the SM preserves periodicity in $y$-direction. One may notice that disorder leads to an enhancement of $W^{\rm SC}$, as shown in Fig.~\ref{fig:delta_dis}~(a). As $U_D$ increases more
SM's subbands couple to the SC's subbands with comparable strength and therefore changes of  $V_g$  that ``push'' different
SM's subbands to the Fermi level do not cause sudden jumps of $W^{\rm SC}$ in contrast to the clean case.
The behavior of $W^{\rm SC}$ versus $V_g$ correlates with the dependence of \deltaind on $V_g$, as shown in Fig.~\ref{fig:delta_dis}~(b). 
Similar to the clean case (see Fig.~\ref{fig:fig7}) there is a one-to-one correspondence between the weight in SC and the induced gap. As follows from Fig.~\ref{fig:delta_dis} the disorder in the SC increases the range of values of $V_g$
for which SC-SM heterostructure is in the strong tunneling regime which agrees qualitatively 
with the recent experiments~\cite{Sole2017, deMoor2018}.

\begin{figure}[t]
\includegraphics[width=0.95\columnwidth]{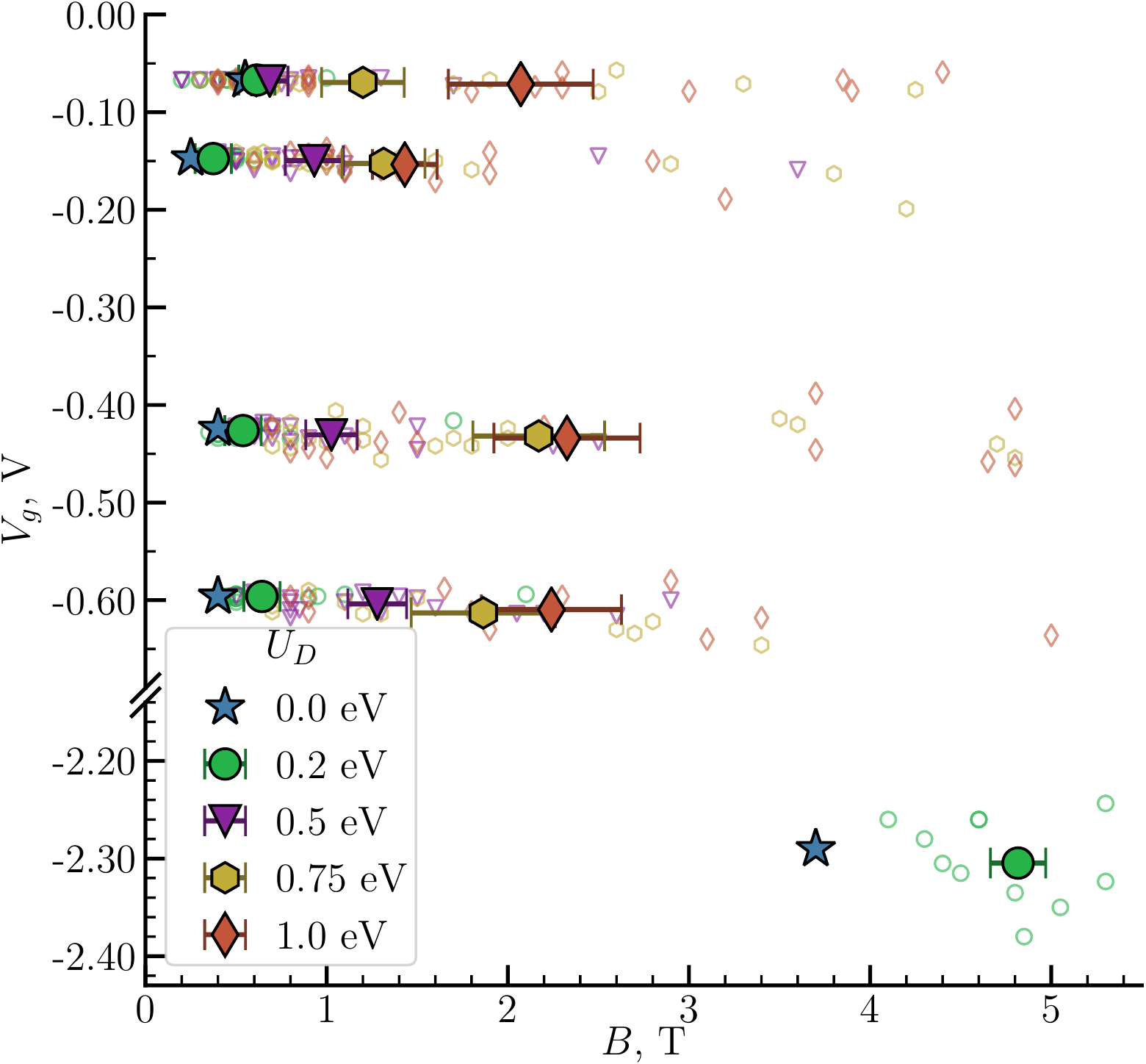}
\caption
 {
     (Color online) Topological phase phase diagram in the presence of disorder. Plotted are the positions of the local minima $B^*_c$ of critical magnetic fields in $B, V_g$ plane. Empty dots corresponds to separate disorder realizations, filled symbols are disorder averages with the corresponding error-bars. Increasing disorder strength leads to larger critical fields. 
}
\label{fig:pd_dis}
\end{figure}

As discussed in the previous sections a large $W^{\rm SC}$ implies not only a large \deltaind but also a reduced g-factor. Since the presence of disorder increases $W^{\rm SC}$ (all other parameters being equal), the reduction of the g-factor is also enhanced.
From Eq.~\ceq{eq:pd2}, we see that both the increase of \deltaind and the reduction of $|g|$ will cause an increase of the minimal critical 
magnetic field for the topological phase transition. Considering
that the disorder in the SC does not affect directly the SM's subbands, in particular their energy at $k_x=0$, we conclude that 
the dominant effect of the disorder in the SC on the topological phase diagram is to cause an increase of $B_c$. 
Therefore, in the presence of disorder the topological regions of Fig.~\ref{fig:pd}~(a) move towards larger values of $B$.
The shift in $V_g$ of the boundaries of the topological phase appears to be negligible.

To study this more quantitatively, we determine the minimal critical field for a topological ``dome'' as
previously shown in Fig.~\ref{fig:pd}; we denote this minimal field as $B^*_c$.
Figure~\ref{fig:pd_dis} shows $B^*_c$ for different values of $U_D$.
The open symbols in Fig.~\ref{fig:pd_dis} represent the values of $B^*_c$ for different disorder realizations whereas the solid symbols correspond to the values of $B^*_c$ averaged over 25 disorder realizations. 
These results clearly demonstrate that as $U_D$ increases $B^*_c$ increases as well, mostly due to a decrease of $g$, and results in a reduction of the area of the topological regions in the $(B,V_g)$ plane.
For very large disorder strengths and negative gate voltages, the effective g-factor becomes so small so that $B^*_c$ becomes larger than the critical field of Al.  

We conclude this section by summarizing that the main effect of disorder in SC is to increase the coupling between SM and SC, resulting in larger induced gap and critical fields to cross the topological phase transition.

\section{Summary and conclusions}\label{sec:conclusions}

We have studied properties of SM-SC nanowires in the presence of strong external electric fields. Our method is based on self-consistent Schr\"odinger-Poisson calculations which treat the semiconductor and the superconductor on equal footing. This approach allows one to take into account several semiconductor subbands which, we believe, are present in current experimental devices. We find that the treatment of the SM and SC at the same level is necessary to describe the strong-coupling regime characteristic to the high-quality epitaxial nanowires~\cite{Krogstrup2015}. Such hybrid nanowires are very promising for the topological quantum computing applications as they exhibit large proximity-induced gaps and very low subgap conductance~\cite{Chang14, Deng2016, Zhang2017}. 

One of the most important results of our work is to provide an insight regarding the necessary conditions for achieving the strong-coupling regime in proximitized nanowires. We find that one of the key ingredients is the presence of an accumulation layer at the interface between the SM and the SC. The presence of an accumulation layer implies a strong confinement of the semiconductor wave function close to the SM-SC interface. Without such confinement, the significant mismatch between the Fermi velocities of SM and SC would significantly reduce the induced gap. This conclusion has recently been supported by angle-resolved photoemission spectroscopy experiments that have shown that in epitaxial InAs/Al systems the band offset $W$ is negative and therefore an electron accumulation layer is present.

We have investigated the effect of an external electric field which can be used to modify the confining potential and, thus, modify properties of electronic states in SM-SC devices. We find that external electric field can be used to change the number of subbands in the semiconductor, tunneling rate, induced gap, magnitude of the effective \gf factor, coherence length $\xi$ etc. Our results show that the relation between \vg and the quantities characterizing the electronic state of SM-SC quasi-1D nanowires is not trivial. The understanding of the interplay of \vg, number of subbands, and electronic properties is one
of the most important results of our work.

We have obtained the topological phase diagram as a function of the gate voltage and magnetic field $B$.
Previous works calculated the topological phase diagram in terms of phenomenological parameters such as effective chemical potential. Our work is the first to present a phase diagram in terms of \vg, the experimentally relevant and tunable quantity, instead of the chemical potential. We find that in the strong coupling regime the renormalization of the \gf factor due to the strong
hybridization between the SM's and the SC's states can significantly modify the topological phase diagram.
For typical setups, the \gf of the SC is smaller (in absolute value) than the SM's \gf factor, and so the strong hybridization reduces the \gf factor causing a decrease in the $(V_g, B)$ plane of the region where the system is in the topological phase.

Finally, we took into account effect of disorder. There is a large body of papers investigating effect of the disorder in the semiconductor~\cite{Motrunich01, Brouwer11a, Stanescu2011, Akhmerov2011, Potter12, Lobos12, Adagideli14, Hegde2016} and at the interface~\cite{Takei2013} concluding that disorder leads to the subgap density of states (i.e. states below the induced gap). However, given the observation of a very small subgap density of states in recent experiments on high-quality proximitized nanowires~\cite{Chang14, Deng2016, Zhang2017,Nichele2017}, we believe that the semiconductor, as well as SC-SM interface, are quite clean. The situation with Al is less clear since the presence of the native oxide covering Al may lead to significant impurity scattering. The effect of the disorder in the superconductor that it relaxes the constraint on momentum conservation and leads to the enhancement of the induced SC gap and critical field. Once again the optimization of the tunneling rate between SM-SC is very important~\cite{Lutchyn2012,DongLiu2017}. 

Our work has important implications for current and future experiments aiming to realize Majorana-based topological qubits using SM-SC heterostructures as it allows one to optimize Majorana devices by tuning key parameters, $\Delta_{\rm}$, \gf, and $\xi$ with gates. Our results show that in the strong coupling the renormalization of \gf can be quite significant increasing the minimal magnetic field necessary to drive the system into the topological phase. Thus, there is a sweet spot, and it is beneficial to operate in the intermediate coupling regime. This is critical information to design experiments aimed at realizing MZMs.  

Finally, we discuss some limitations of our model. First, the disorder is two-dimensional, leading to a qualitatively similar picture as in the disorder-free case. Impurities in the SC may induce subgap states~\cite{Lutchyn2012, Hui2015, Cole2016, DongLiu2017} that can be captured in a fully three-dimensional simulation. Another limitation of our model is the lack of orbital effects due to the magnetic field. Due to the strong geometry dependence of the orbital effect~\cite{Winkler2017, Wojcik2018}, however, a careful treatment of it needs to go beyond the slab model discussed here \cite{deMoor2018, Winkler_in_prep}.

Finally,  we emphasize that, although in this work we focused on InAs/Al hybrid nanowires, the Schr\"odinger-Poisson approach proposed in this work can be used to study other heterostructures such as InSb/Al nanowires, two-dimensional SM-SC heterostructure and the quasi 1D channels created by electrostatic confinement in such structures.  

\textit{Note added}.
After our arXiv posting, three other preprints on a similar topic appeared~\cite{Woods2018,Mikkelsen2018,Reeg2018}.
These manuscripts, as ours, present numerical approaches aimed at a more quantitative description of semiconductor-superconductor heterostructures. 
Reference \onlinecite{Woods2018} studies the electrostatic environment of nanowires in the weak coupling regime. Ref. \onlinecite{Mikkelsen2018} discusses the electrostatic environment in the presence of metallic Al and focuses on the normal-state band structure of the quasi-one-dimensional heterostructure. Reference \onlinecite{Reeg2018} studies the renormalization of the semiconductor band structure by the proximity to the bulk superconductor in the strong coupling regime neglecting electrostatic effects.

\section{Acknowledgments}
We are grateful to Mingtang Deng, Karsten Flensberg, John Gamble, Jan Gukelberger, Peter Krogstrup, Bas Nijholt, Saulius Vaitiek\'enas,  Adriaan Vuik, Michael Wimmer, and Hao Zhang for stimulating discussions. ER acknowledges support from NSF-DMR-1455233, ONR-N00014-16-1-3158, and ARO-W911NF-16-1-0387.

\bibliography{toporefs6}

\end{document}